\documentclass[lettersize,journal]{IEEEtran}
\usepackage{amsmath,amsfonts}
\usepackage{dsfont}
\usepackage{algorithmic}
\usepackage{algorithm}
\usepackage{array}
\usepackage{textcomp}
\usepackage{stfloats}
\usepackage{url}
\usepackage{verbatim}
\usepackage{graphicx}
\usepackage{cite}
\usepackage{xcolor}
\usepackage[font=small]{caption}
\usepackage{subcaption}
\usepackage{tikz}
\usepackage[normalem]{ulem}
\usepackage{amsfonts, amsthm, amssymb}
\usepackage{multirow,verbatim}
\usepackage{soul}

\newcommand{\id}{\mathds{1}}
\newcommand{\crd}{\mathtt{Crd}}
\newcommand{\img}{\mathtt{Img}}
\newcommand{\ldr}{\mathtt{LiD}}
\newcommand{\loss}{\mathcal{L}}

\newcommand{\env}{\mathcal{S}}

\newcommand{\twin}{\mathcal{T}}
\newcommand{\labeled}{l}
\newcommand{\unlabeled}{u}
\newcommand{\numberseen}{\mathtt{L}}
\newcommand{\numberunseen}{\mathtt{U}}
\newcommand{\multiverse}{\mathcal{M}}

\newcommand{\beam}{\mathcal{B}}

\newcommand{\power}{\mathtt{p}}

\newcommand{\cost}{\mathcal{C}}

\newcommand{\reflect}{\rho}
\newcommand{\transpath}{\iota}
\newcommand{\diffract}{\zeta}
\newcommand{\antR}{\varsigma}
\newcommand{\first}{\text{Baseline}}
\newcommand{\second}{\text{1-Reflection}}
\newcommand{\third}{\text{3-Reflection}}
\newcommand{\WI}{\text{WI}}
\newcommand{\MV}{\text{Multiverse}}

\usepackage{color, colortbl}
\definecolor{lightgray}{RGB}{211,211,211}
\definecolor{CyanT1}{RGB}{230,255,255}
\definecolor{CyanT2}{RGB}{210,255,255}
\definecolor{CyanT3}{RGB}{190,255,255}

\newcolumntype{L}[1]{>{\raggedright\let\newline\\\arraybackslash\hspace{0pt}}m{#1}}
\newcolumntype{C}[1]{>{\centering\let\newline\\\arraybackslash\hspace{0pt}}m{#1}}
\newcolumntype{R}[1]{>{\raggedleft\let\newline\\\arraybackslash\hspace{0pt}}m{#1}}

\begin{document}
\title{Multiverse at the Edge: Interacting Real World and Digital Twins for Wireless Beamforming}
\author{
\IEEEauthorblockN{Batool Salehi\IEEEauthorrefmark{1}\,\IEEEauthorrefmark{3}, Utku Demir\IEEEauthorrefmark{1}\,\IEEEauthorrefmark{2},  Debashri Roy\IEEEauthorrefmark{2}, Suyash Pradhan\IEEEauthorrefmark{2},\\ Jennifer Dy\IEEEauthorrefmark{3}, Stratis Ioannidis\IEEEauthorrefmark{3}, Kaushik Chowdhury\IEEEauthorrefmark{3}} \\ \textit{Institute for the Wireless Internet of Things}, Northeastern University, Boston, MA, USA \\ \IEEEauthorblockN{{\IEEEauthorrefmark{2}\{u.demir, d.roy, pradhan.suy\}@northeastern.edu}, \IEEEauthorrefmark{3}\{bsalehihikouei, jdy, ioannidis, krc\}@ece.neu.edu} \\ 
\thanks{*Batool Salehi and Utku Demir have equally contributed to this work.}
}

\maketitle
\thispagestyle{plain}
\pagestyle{plain}
\begin{abstract}
Creating a digital world that closely mimics the real world with its many complex interactions and outcomes is possible today through advanced emulation software and ubiquitous computing power. Such a software-based emulation of an entity that exists in the real world is called a `digital twin'. In this paper, we consider a twin of a wireless millimeter-wave band radio that is mounted on a vehicle and show how it speeds up directional beam selection in mobile environments. To achieve this, we go beyond instantiating a single twin and propose the `$\MV$' paradigm, with several possible digital twins attempting to capture the real world at different levels of fidelity. Towards this goal, this paper describes (i) a decision strategy at the vehicle that determines which twin must be used given the computational and latency limitations, and (ii) a self-learning scheme that uses the $\MV$-guided beam outcomes to enhance DL-based decision-making in the real world over time. Our work is distinguished from prior works as follows: First, we use a publicly available RF dataset collected from an autonomous car for creating different twins. Second, we present a framework with continuous interaction between the real world and $\MV$ of twins at the edge, as opposed to a one-time emulation that is completed prior to actual deployment. Results reveal that $\MV$ offers up to $79.43\%$ and $85.22\%$ top-$10$ beam selection accuracy for LOS and NLOS scenarios, respectively. Moreover, we observe $52.72-85.07\%$ improvement in beam selection time compared to 802.11ad standard. 
\end{abstract}


\section{Introduction}
\label{sec:Sec1_introduction}
A {\em digital twin} is a software-based emulation of a physical entity that captures its real world properties and interactions in the environment in which it operates~\cite{nguyen2021digital}. Thus, it allows tracking the state-changes of the real entity over time and also studying the impact of any configuration settings in a safe, digital environment. 
In wireless communication domain, digital twins are used for emulating 5G networks, modeling wireless channels, and validation and optimization~\cite{khan2022digital_opp}. However, recent examples of such twins rely on one specific realization of the twin in the digital domain~\cite{khan2022digital}.
The rich diversity in edge computing-enabled wireless network infrastructures raises an intriguing possibility: {\em What if there are `multiple such twins’, with each twin capturing the real world wireless channel and signal propagation with a different level of fidelity.}  In this paper, we introduce the novel paradigm of the {\em $\MV$ of twins}, where the goal is to define an analytical method that allows the system to select one of multiple candidate twins, according to the computation and latency constraints. 

\begin{figure}[t!]
\centering
  \includegraphics[width=\linewidth]{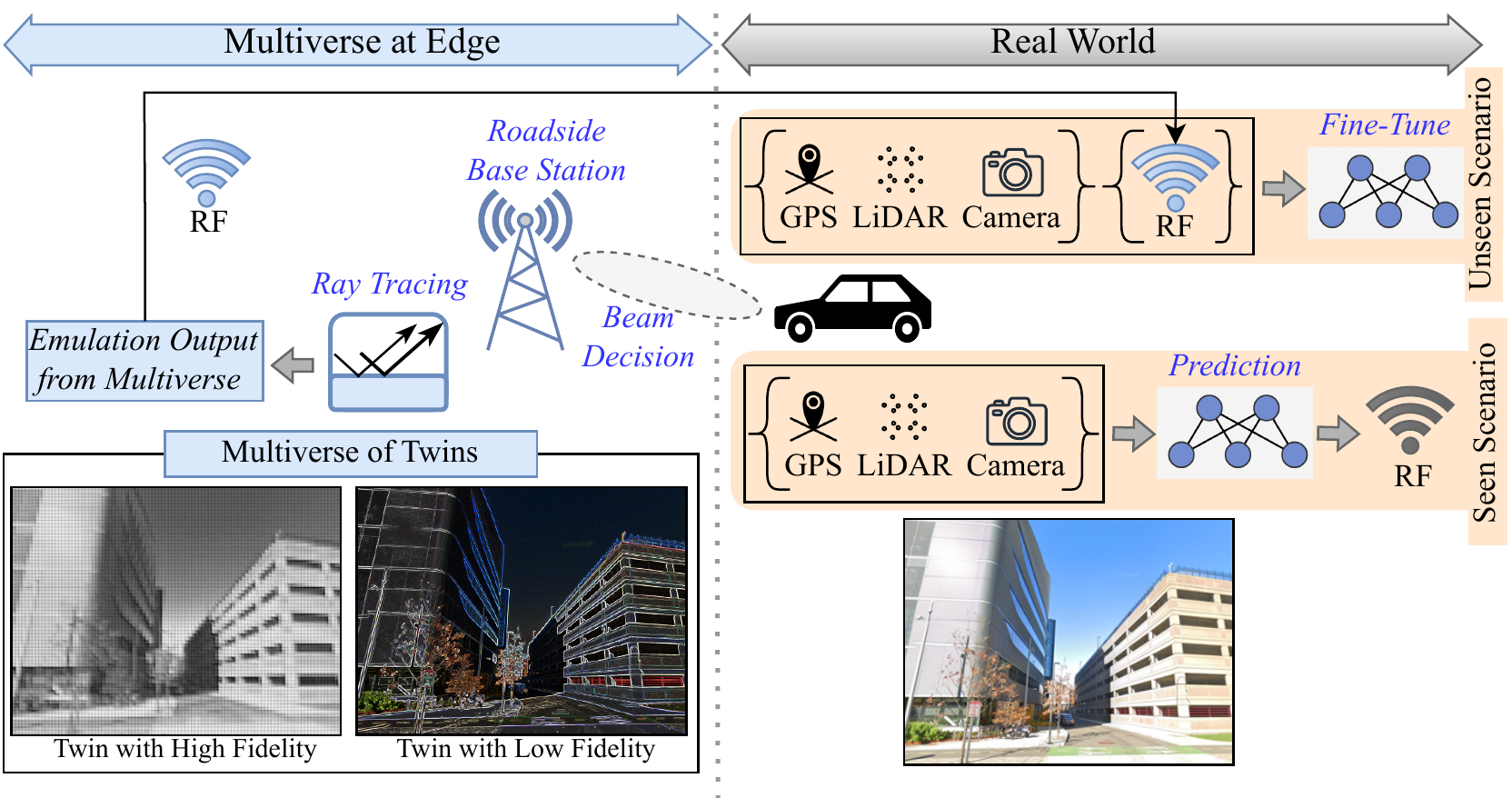}
  \caption {Overview of the proposed framework. The sensors on the vehicle capture the state of the real world. They may use this local information and DL models to choose a beam directly or invoke a twin from the $\MV$. Each twin offers a distinct fidelity of emulation through ray tracing and incurs a computation cost for delivering the beam selection results back to the vehicle.} 
  \vspace*{-10pt}
  \label{fig:intro}
\end{figure}

\subsection{Deep Learning for Beamforming in Seen Scenarios}
As shown in  Fig.~\ref{fig:intro}, our use case involves selecting one of several directional beams in the millimeter-wave (mmWave) band to establish connectivity between an autonomous car and a roadside base station (BS). Vehicle-mounted sensors, e.g., camera, GPS, and LiDAR are used to obtain contextual information about the environment. In recent works~\cite{salehi2022deep,salehi2022flash}, we have studied  deep learning (DL) models with convolutional neural networks (CNNs) to fuse available multimodal sensor data and predict the best beam at the BS. This considerably shortens the time by {\color{black}$52.75\%$ and $95\%$} compared to the conventional exhaustive-sweeping approach, defined in the 802.11ad and 5G-NR standards, respectively. However, a challenge arises when the DL model needs to perform in test environments that it has not encountered previously during training~\cite{pang2021deep}~(an unseen scenario), a new street or temporary obstruction, for example. Such situations result in unpredictable wireless propagation conditions~($\sim 73\%$ drop in prediction accuracy according to our experiments) that cannot be fully characterized without new RF information.

\subsection{Multiverse at Edge for Beamforming in Unseen Scenarios}
We propose the $\MV$ a software-based paradigm that runs ray tracing to emulate the RF propagation patterns, in unseen environments. This has the potential to replace--and correct--erroneous DL predictions, while still avoiding the latency of exhaustive sweep, defined by the standards. In the $\MV$, several twins, each offering a different levels of fidelity and associated computation cost, coexists and ray tracing emulation is performed for all of them, in each unseen scenario. The emulation cost of each twin increases with its level of fidelity. Thus, the accuracy of emulating `reality' is determined by the available computational resources at the edge, such as whether having access to a CPU unit, commercial grade GPUs, or a GPU farm. On the other hand, depending on wireless latency constraints, choosing a twin with lower complexity may be favored. Our framework is apt for exploring this accuracy-latency trade-off by selecting a twin from the `$\MV$' and set of Top-$K$ beam recommendations within the selected twin, according to the computation and latency constraints. Moreover, the ray tracing outputs from the $\MV$ can also generate a labeled data point for continuous learning of the DL models at the vehicle. Ultimately, the prediction of the twin must be timely~(i.e., the vehicle can use its prediction before it speeds away) and accurate~(i.e., close to exhaustive search). Taken together, the main challenges involve (i) detecting seen and unseen scenarios locally at the vehicle, (ii) choosing one of the twins from the $\MV$ in an unseen scenario based on the computation and latency constraints, (iii) ensuring the complete cycle of query-compute-response can be completed in less time than performing exhaustive search locally.

\subsection{Overview of the Proposed Framework}
In our proposed framework, the vehicle first detects if it encountered a seen or unseen scenario. In a seen scenario, the vehicle predicts the beam locally using DL-based method and there is no need to invoke any twin from the $\MV$. In an unseen scenario, e.g., a blockage of certain dimensions that was not included in the training data for DL-based method, we use the $\MV$ for beamforming instead. Upon triggering the $\MV$, if the unseen environment is not included in the $\MV$ of twins, the vehicle relays the sensor data to setup a digital replica in the ray tracing tool Wireless InSite~\cite{WI_webpage} that models the geographical location of the car, dimension and placement of the obstacles, twin-specific antenna models, and multipath emulation settings. At the end of this step, lookup tables are generated and added to the $\MV$ of twins to be used by upcoming vehicles later. These look up tables include information about the locations of the receiver~(Rx) and signal-to-noise-ratio~(SNR) of all beams obtained by ray tracing. If the new unseen environment is already included in the $\MV$, the pre-generated look up tables for all twins are shared with the vehicles in the downlink. In the next step, a decision strategy at the vehicle identifies (a) which of the twins from the $\MV$ must be used, and (b) what is the optimum subset of beams for the selected twin, according to the latency and computation constraints. The candidate top-$K$ beams are then swept by the receiver to identify the optimum beam and start the transmission. Finally, the ray tracing outputs from the $\MV$ are paired with the local sensor data to fine-tune the DL model at the vehicle for the future. 


\subsection{Summary of Contributions.} 
Our main contributions are as follows:
\begin{itemize}
    \item We propose the $\MV$ paradigm, where different twins emulate the real world with varying levels of fidelity. We propose an optimization algorithm that takes into account the fidelity of the twins and user-defined latency and computation constraints to automatically select the optimum twin from the $\MV$ and associated top-$K$ beams in a {\em case-by-case} basis. This is locally executed on the vehicle with complexity $O(N)$ with N being the number of twins in the $\MV$.
    \item We propose to leverages the ray tracing outputs from the $\MV$ to generalize the real world DL models to unseen environments, in actual deployment conditions. Thus, the labels from the $\MV$ and real world sensor data are paired to fine-tune the local DL model. We observe that the top-$10$ accuracy with fine-tuning is bounded by top-$1$ and top-$10$ accuracy of the $\MV$.

    \item We rigorously evaluate the $\MV$ paradigm and how well a twin's predictions match with the ground-truth, using a publicly available real world dataset~\cite{flash_dataset} for vehicle-to-infrastructure~(V2I) communication. Our results reveal that the $\MV$ offers up to $79.43\%$ and $85.22\%$ top-$10$ beam selection accuracy for Line-of-Sight~(LOS) and non-Line-of-Sight~(NLOS) scenarios, respectively. Finally, we demonstrate that the $\MV$ decreases the beam selection overhead by $52.72-85.07\%$ compared to exhaustive search method proposed by the state-of-the-art 802.11ad standard.

    \item We pledge to publish the {\em first-of-its-kind} dataset and simulation code for the $\MV$ with precise maps, experimentally measured antenna patterns, and building materials using the {\em Wireless InSite}~\cite{WI_webpage} software. We provide APIs to interface the Wireless InSite software framework with our previously released FLASH dataset~\cite{flash_dataset} obtained from a sensor-equipped Lincoln MKZ autonomous car.
\end{itemize}

\section{Related Work}
\label{sec:related_work}
As this paper describes a wireless use case, we limit the related work discussion on digital twins within this domain. 

\subsection{Digital Twin for Optimization}
Dong {\em et al.}~\cite{dong2019deep} train a DL model for base station and user association to optimize the energy consumption. The DL model is periodically updated in the digital twin, using the information flow from the real world. 
Similarly, Lu {\em et al.}~\cite{lu2021adaptive} formulates how to set up a digital twin in a mobile network via reinforcement and transfer learning in order to ensure connectivity with low computation latency.
Sun {\em et al.}~\cite{9931961} propose a lightweight digital twin framework within a network of UAVs, which performs energy-efficient UAV placement with the objective of establishing connectivity with ground units in emergency situations. 
Boas {\em et al.}~\cite{9897088} propose to use a two-stage deep learning scheme and digital twin for channel state information (CSI) estimation. The authors use untrained neural networks (UNNs) to learn the CSI,  which is then fed into a conditional generative adversarial networks (cGAN) to create the corresponding digital world. After this step, their framework only needs the node locations to generate CSI.


\subsection{Digital Twin for Beamforming}
Li {\em et al.}~\cite{9593108} use digital twin to generate training data for channel estimation, which is important in beamforming. The emulated 5G complaint data is used to train a deep generative model that predicts the channel using only the precoder matrix indicator (PMI).
Zeulin {\em et al.}~\cite{9773088} use digital twin to generate training data~(channel map) through a ray tracing simulator in mmWave band. They then train a model to a predict subset of candidate AoA/AoD options. The success measure of their Machine Learning~(ML) model is the probability of capturing half-power beamwidth and error in AoA. Cui {\em et al.}~\cite{cui2023digital} utilize digital twins for beamforming with the assitance of Reconfigurable Intelligence Surfaces (RIS). They jointly optimize user association and power allocation by running a deep reinforcement learning~(DRL) at the twin. The success of their eventual goal, i.e. maximizing sum-rate of uplink in the user-centric cell-free systems, is demonstrated in simulation.

\noindent{\bf Novelty of the Multiverse:} In summary, the state-of-the-art work concerning digital twin in wireless applications is based on simulation alone. Also, to the best of our knowledge, current wireless digital twin models are applicable for stationary or pedestrian-speed mobility. Our work distinguishes itself over prior work by: (a) using real data for validation, (b) including vehicular-mobility components, (c) offering a $\MV$ of twins to choose from, (d) proposing a pioneering work (to the best of our knowledge) on interactive digital twin based beam selection in mmWave band. 
Also, unlike prior work, we pledge to release datasets and software APIs for rigorous validation by other researchers. 

\section{System Architecture}
\label{sec:system_architecture}
In this section, we first review classical beam initialization and formally present the beam selection problem. We then introduce the system architecture in our framework that exploits the DL-based method for seen and $\MV$ for unseen scenarios~(see Fig.~\ref{fig:intro}). We summarize the notations in Table~\ref{tab:notation}.

\begin{table}[t!]
    \centering
    \resizebox{\linewidth}{!}{\color{black}{\begin{tabular}{|p{2.8cm}|p{6cm}|}
    \hline
        {\bf Notation} & {\bf  Description} \\
        \hline
        \hline
        $C_{Tx}$ & Codebook of transmitter with $\beam$ beams\\
        $\power_{t_{b_i}}$ & Received signal strength for beam $t_{b}\in C_{Tx}$\\
        $\env_\labeled$ & Seen scenario  \\
        $\env_\unlabeled$ & Uneen scenario  \\
        ${X}_{l,j}^{\ldr}, {X}_{l,j}^{\img}, {X}_{l,j}^{\crd},{Y}_{l,j}$ & LiDAR, image, and GPS data and ground-truth for\\
        & sample $j$ and $l^\text{th}$ seen scenario\\
        ${\mathbb{X}}_{l,j}^{\ldr}, {\mathbb{X}}_{l,j}^{\img}$, ${\mathbb{X}}_{l,j}^{\crd}$ & Test samples from LiDAR, image and coordinate for sample $j$ and $l^\text{th}$ seen scenario\\
        $n_\labeled$ & Number of samples in $l^\text{th}$ seen scenario \\
        $f_{\theta}$ & Learning model\\
        $\mathcal{L}({\theta}; \env_\labeled)$ & Loss function for seen scenario $\env_\labeled$\\
        $t^{*}_{l,j}$ & Best beam for sample $j$ and $l^\text{th}$ seen scenario\\ 
        $\multiverse_u$ & Multiverse for unseen scenario $u$ \\
        $\twin_{\unlabeled,i}$ & $i^\text{th}$ twin for unseen scenario $u$ \\
        $map_{u}$ & Imported OpenStreetMap for unseen scenario $u$\\
        $\mathrm{O}_{u}$ & Present structures or obstacles for unseen scenario $u$\\
        $C^{Tx}_{i}$ & Transmitter codebook for twin $\twin_{\unlabeled,i}$~($|C^{Tx}_{i}|=\beam_i$)\\
        $\reflect_{i}$ & Number of allowed reflections for $i^\text{th}$ twin\\
        $N^{obs}_{u}$ & Number of obstacles in unseen scenario $\env_\unlabeled$\\
        $L_{\twin_{\unlabeled,i}}$ & Set of location to perform ray tracing \\
        &for twin $\twin_{\unlabeled,i}$~($|L_{\twin_{\unlabeled,i}}|=n_{\twin_{\unlabeled,i}}$)\\
        $l_{j}$ & A single point from $L_{\twin_{\unlabeled,i}}$\\
        $N_{j}^{m,i}$ & Number of total propagation rays delivered at Rx at location $l_{j}$ for the Tx beam $t_m$ and twin $\twin_{\unlabeled,i}$\\
        $\power_{j}^{m,i}$ & Power at location $l_{j}$ for Tx beam  $t_m$ in twin $\twin_{\unlabeled,i}$\\
        $\mathcal{N}$ & Noise power \\
        $\texttt{SNR}_{j}^{m,i}$ & SNR at location $l_{j}$ for Tx beam  $t_m$ in twin $\twin_{\unlabeled,i}$\\
        $\cost^{\twin_{\unlabeled,i}}_{Comp}$ & Computation cost of of twin $\twin_{\unlabeled,i}$ including cost of \\ &generating map~($\cost^{\twin_{\unlabeled,i}}_{map}$) and look up tables~($\cost^{\twin_{\unlabeled,i}}_{lookup}$)\\
        $\transpath$ & Number of transmissions through walls\\
        $\diffract$ & Number of allowed diffraction\\
        $w$ &  Number of reflective surfaces reached by the rays emitted from the Tx \\
        $LT (\twin_{\unlabeled,i})$ & Look up table for twin $\twin_{\unlabeled,i}$\\
        $\cost^{\texttt{5G-NR}}(K)$ & Communication cost in 5G-NR for $K$ beams\\
        $p(.)$ & Probability of inclusion\\
        $\gamma(.)$ & Outlier detection algorithm \\
        \hline
    \end{tabular}}}
        \caption{Notation Summary}
    \label{tab:notation}
\end{table}
\subsection{Traditional Beam Initialization} 
The 802.11ad and 5G-NR standards use exhaustive search for beamforming wherein the transmitter (Tx) sends out probe frames in all beams sequentially and the receiver (Rx) listens to these frames with a quasi-omni-directional antenna setting~\cite{nitsche2014ieee}. This process is then repeated with the Tx-Rx roles reversed. Assume the Tx has a pre-defined codebook $C^{Tx}=\{t_1,\cdots,t_{\beam}\}$, consisting of $\beam$ elements. After transmitting $\beam$ probe frames, the beam with the maximum signal strength is chosen as optimal. Formally, the optimal beam at Tx is calculated as: 
\begin{equation}
   t^* = \underset{1\leq m \leq \beam}{\arg\max}~\power_{t_{m}},
   \label{eq:best_sector}
 \end{equation}
where $\power_{t_{m}}$ is the received signal strength at the Rx when the transmitter uses beam $t_{m}$. The exhaustive search methods are slow and particularly impractical in a V2X scenarios, where the optimal beam changes rapidly, due to mobility.

\subsection{DL-based Prediction at Vehicle and Multiverse at Edge}

The RF propagation pattern in the mmWave band is affected by factors such as the positioning of the transmitter and receiver, atmospheric features, and presence of obstacles specially the ones that block the LOS. As a result, in dynamic scenarios such as vehicular networks, the RF profiles continuously change over time which makes the DL-based methods prone to faulty predictions in unseen scenarios. 
Thus, we propose an interactive $\MV$-based solution, where the vehicle: (a) recognizes the unseen scenarios locally; (b) invokes a twin in the $\MV$ for beam selection; (c) uses emulation outputs for fine-tuning the local DL models. Our proposed frameworks runs by coordination between the DL-based beam selection and $\MV$ and has three main thrusts as follows:

\begin{itemize}
    \item \textbf{Beam Selection using DL-based framework:}
    While encountering a seen scenarios, we use the DL-based beam selection that exploits the local sensor data at the vehicle to predict the optimum beam (see Sec.~\ref{sec:ml_beam_selection}).
    \item \textbf{Beam Selection using Multiverse at Edge:} We use the $\MV$ including twins with different levels of fidelity and associated computation cost at the edge, for unseen scenarios. We first create the $\MV$ offline by emulating the propagation patterns using a ray tracing tool, where we consider different number of reflections and custom beam patterns. After this offline creation step, a lookup table is generated for each twin that contains the SNR of the beams and twin querying cost. The vehicles use these look up tables to choose the optimum twin from the $\MV$ and top-$K$ beams locally, while considering the communication and computation constraints~(see Sec.~\ref{sec:multiverse_beam_selection}).
        
     \item \textbf{Fine-Tuning with Multiverse Ground-Truth:} We incorporate the the emulated beams from the $\MV$ to label the sensor data at the vehicle. Thus, we fine-tune the local model to account for previously unseen scenarios in the future~(see Sec.~\ref{subsec:realworld_augmentation}).
\end{itemize}

\section{Multiverse-based Beam Selection}
\label{sec:proposed_method}
\begin{figure*}[t]
    \centering
    \includegraphics[width=\textwidth]{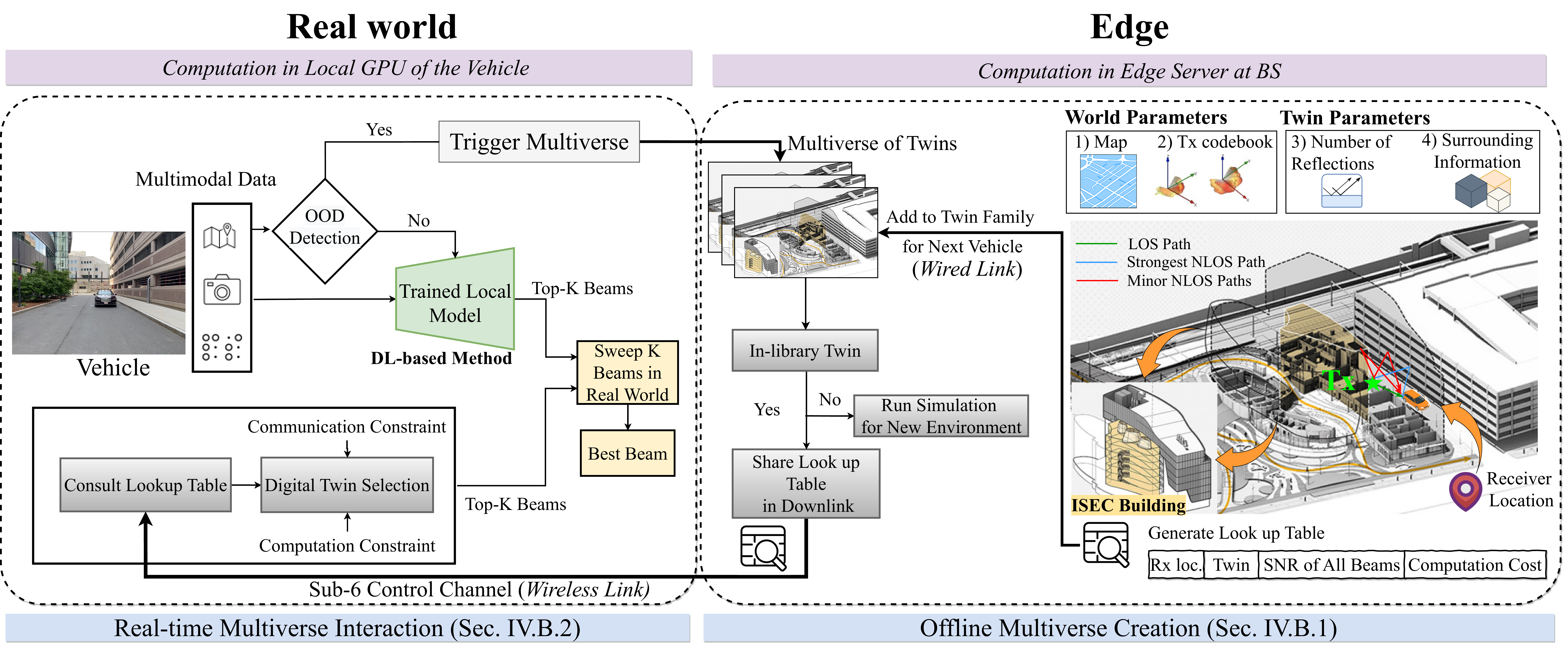}
    \caption{The system model of the proposed $\MV$-based beam selection framework. In the offline $\MV$ creation step, we run ray tracing emulation to obtain the beam profiles for each twin. We report the ray tracing output as a look up table including the, Rx location index, Twin index,  SNR values for all beams of twin $\twin_{\unlabeled,i}$ over all locations $l_{j}\in L_{\twin_{\unlabeled,i}}$, and computation cost $\cost^{\twin_{\unlabeled,i}}_{Comp}$ associated to twin $\twin_{\unlabeled,i}$. The lookup table is retrieved from the $\MV$~(at the edge) by the vehicle. In the real-time $\MV$ interaction, the vehicles solve Eq.~\ref{eq:opt_twin_selection} to identify the optimum twin and subset of $K$ beams from the selected twin.}
    \label{fig:system_model}
\end{figure*}

In this section, we introduce our mmWave beam selection framework, consisting of the DL-based prediction at the vehicle and $\MV$ at edge. We denote the set of seen~(i.e. labeled) and unseen~(i.e. unlabeled) scenarios by $\{\env_\labeled\}_{l=1}^{\numberseen}$ and $\{\env_\unlabeled\}_{u=1}^{\numberunseen}$, respectively. Here, $\env_\labeled$ and $\env_\unlabeled$ are samples from a total of $\numberseen$ and $\numberunseen$ seen and unseen scenarios, respectively. When a vehicle comes within a range of $\MV$-enabled mmWave base station, it downloads: (a) a trained DL model that utilizes multimodal sensor data for seen scenario $\env_\labeled$, and (b) the lookup tables from the $\MV$ for unseen scenario $\env_\unlabeled$. Fig.~\ref{fig:system_model} shows the interactions of real world and $\MV$ in our proposed framework. 

\subsection{Beam Selection using DL-based framework}
\label{sec:ml_beam_selection}
To avoid the costly exhaustive search in traditional beam selection, one approach is using DL models that predict the best beam using non-RF multimodal data, such as LiDAR, camera images, and GPS coordinates~\cite{salehi2022flash, salehi2022deep}. In this method, a training set is available prior to deployment for seen environments $\{\env_\labeled\}_{l=1}^{\numberseen}$. The training data corresponding to the $l^{\text{th}}$ seen scenario includes the multimodal sensor data as well as the RF ground-truth $\{({X}_{l,j}^{\ldr}, {X}_{l,j}^{\img}, {X}_{l,j}^{\crd}),{Y}_{l,j}\}_{j=1}^{n_l}$. Here, ${X}_{l,j}^{\ldr}, {X}_{l,j}^{\img}, {X}_{l,j}^{\crd}$ are LiDAR, image, and GPS coordinate samples and ${Y}_{l,j}\in \{0,1\}^{\beam}$ is the corresponding label for the sample $j$ and $l^\text{th}$ seen scenario. Moreover, $n_l$ is the total of training samples for seen scenario $\env_\labeled$.

The learning model $f_{\theta}(.)$ is a function parameterized by $\theta$, i.e., a neural network with weights $\theta$. The empirical loss of the model parameters ${\theta}$ on dataset for seen scenario $\env_l$ is defined as $\loss({\theta}; \env_l) := \frac{1}{n_l}\sum_{j=1}^{n_l} [\ell(f_{\theta}({X}_{l,j}^{\ldr}, {X}_{l,j}^{\img}, {X}_{l,j}^{\crd}), Y_{l,j})]$, where $\ell$ is a loss function measuring the discrepancy between predicted and true labels, cross entropy as an instance.
The standard DL training approach finds a model that minimizes the loss across all of the training samples by solving:
    $\underset{\theta} {\min} \frac{1}{U} \sum_{l=1}^\numberseen n_l \mathcal{L}({\theta}, \env_l),$ 
with $U=\sum_{l=1}^L n_l$. After the model training, the best beam is predicted as: 
\begin{equation}
    t^{*}_{l,j} = \sigma(f_{\theta}({\mathbb{X}}_{l,j}^{\ldr}, {\mathbb{X}}_{l,j}^{\img}, {\mathbb{X}}_{l,j}^{\crd})),
\label{eq:best_beam_pair}
\end{equation}
where $\sigma$ denotes the {\em softmax} activation,  ${\mathbb{X}}_{l,j}^{\ldr}, {\mathbb{X}}_{l,j}^{\img}$, and ${\mathbb{X}}_{l,j}^{\crd}$ are test samples from LiDAR, image and coordinate sensors for sample $j$ and $l^\text{th}$ seen scenario, respectively. 

\subsection{Beam Selection using Multiverse at Edge}
\label{sec:multiverse_beam_selection}
The proposed DL-based beam selection method in Sec.~\ref{sec:ml_beam_selection} exploits the sensor data to predict the optimum beam locally at the vehicles. However, it fails to provide accurate prediction for unseen scenarios $\{\env_\unlabeled\}_{u=1}^{\numberunseen}$. In the case of an unseen scenarios, the $\MV$ is triggered to emulate the beam profiles instead, which operates at the edge. In the $\MV$, a collection of $N$ digital twins are available with different levels of fidelity: $\multiverse_\unlabeled =  \{\twin_{\unlabeled,i}\}_{i=1}^{N}$ for each unseen scenario $\env_\unlabeled$. After running the ray tracing emulation, lookup tables are generated for each twin in the $\MV$, with per-beam SNRs over different locations on the road. These lookup tables are downloaded to the local processing unit of the vehicle, suppose a local GPU and are used for beam selection. In this section, we explain how we create the $\MV$ of twins at the edge and present our strategy to determine the optimum twin from the $\MV$ and set of top-$K$ beams at the vehicle.
\subsubsection{Offline Multiverse Creation}
\label{subsec:multiverse_creation}
The $\MV$ creation for each unseen scenario $\env_\unlabeled$ consists of three steps: (a) twin-world creation~(map geometry), (b) modeling the propagation paths using a ray tracing tool, (c) creation of lookup table. Having a high fidelity twin of the real world RF propagation patterns is complex and time intensive. Thus, we assume that creating the twins is performed offline. Nevertheless, with enough computation resources, it can be completed in near real-time, as studies in this direction are under development~\cite{hoydis2023sionna}.

\noindent\textbf{Twin-world Creation.} In the $\MV$, we consider properties with respect to map precision, transmitter codebook, and mmWave propagation properties. While we consider several key metrics in the $\MV$, our baseline can be extended in the future to twins that also incorporate weather patterns, such as the effect of rain/snow. We initialize the $i^\text{th}$ twin for unseen scenario $\env_\unlabeled$ as $\twin_{\unlabeled,i}= f_{twin}(map_{u},\mathrm{O}_{u}, C^{Tx}_{i}, \reflect_{i})$ in the $\MV$. Here, $map_{u}$ and $\mathrm{O}_{u}$ denote the imported OpenStreetMap~\cite{OpenStreetMap_webpage} and present structures or obstacles for unseen scenario $\env_\unlabeled$. Moreover, $C^{Tx}_{i}$ and $\reflect_{i}$ are transmitter codebook and number of allowed reflections for $i^\text{th}$ twin. In our design, each twin can have a different codebook with $\beam_i$ beams for $i^\text{th}$ twin~($|C^{Tx}_{i}|=\beam_i$). We characterize present objects in the twin with 
$\mathrm{O}_{u} = \{(x_k, y_k, d_{0,k}, d_{1,k}, d_{2,k})\}_{k=1}^{N^{obs}_{u}},$
where $N^{obs}_{u}$ is the number of structures or obstacles that are present in the unseen scenario $\env_\unlabeled$ and are replicated in the $\MV$. Moreover, ($x_k$, $y_k$) are the obstacle coordinates on the map, whereas $d_{0,k}$, $d_{1,k}$, $d_{2,k}$ represent the height, width, and length of the obstacle-$k$.

\noindent\textbf{Modeling the Propagation Paths.} To model the propagation paths in twin $\twin_{\unlabeled,i}$, we place the receiver at $L_{\twin_{\unlabeled,i}}=\{l_{j}\}_{j=1}^{n_{\twin_{\unlabeled,i}}}$ different locations on the road and send out mmWave rays from the BS, where the Tx is located, with $\reflect_i$ reflections for each location. We then perform the propagation study for each beam $t_m\in C^{Tx}_{i}$ at each $l_{j}\in L_{\twin_{\unlabeled,i}}$ location.



We follow X3D~\cite{WI_webpage} for modeling the mmWave propagation path that takes into account the phase information of the rays. We calculate the total received power, i.e. aggregated power from individual rays at the Rx, for the Tx beam $t_m$, twin $\twin_{\unlabeled,i}$, and location $l_{j}$ as:
\begin{equation}
    \power_{j}^{m,i} = \frac{\lambda^2 \beta}{8 \pi \eta_{0}} \left| \sum_{n=1}^{N_{j}^{m,i}} [E_{\theta,n}g_{\theta}(\theta_{n}, \phi_{n}) + E_{\phi,n}g_{\phi}(\theta_{n}, \phi_{n})] \right|^2,
    \label{eq:P_r}
\end{equation}
where $N_{j}^{m,i}$ is the number of total propagation rays delivered at Rx at location $l_{j}$ for the Tx beam $t_m$ and twin $\twin_{\unlabeled,i}$. Here, $\lambda$ is the wavelength, and $\beta$ is the 
ratio of the area under the overlapping frequency spectrum of the transmitted and received signals to the area under the frequency spectrum of the transmitted signal~\cite{WI_references}.
Moreover, $\eta_{0}$ is the impedance of free space (where $\eta_{0}$ = $377\Omega $), $\theta_{n}$ and $\phi_{n}$ represent direction of arrivals, $E_{\theta,n}$ and $E_{\phi,n}$ are the theta and phi components of the electric field of the $n^{th}$ ray at the Rx. Further, $g_{\phi}(\theta, \phi) = \sqrt{|G_{\theta} (\theta, \phi)|}e^{j\psi_{\theta}}$, where $G_{\theta}$ is the theta component of the receiver antenna gain, and $\psi_{\theta}$ is the relative phase of the $\theta$ component of the far zone electric field. Once the received power is determined~(in Watts), the power in $dBm$ is calculated using $\power_{j}^{m,i}(dBm) = 10\log_{10}[\power_{j}^{m,i}(W)] - L_{s}(dBm)$, where $L_{s}$ is any loss in the system other than path loss, such as within cables and electronics at the Rx.
We calculate the SNR values for the beam $t_m\in C^{Tx}_{i}$ at location $l_{j}\in L_{\twin_{\unlabeled,i}}$ as $\texttt{SNR}_{j}^{m,i} = \mathtt{P}^{m,i}_{j} - \mathcal{N}$, where $\mathcal{N}$ is the noise power in $dBm$ at location $l_{j}$ for twin $\twin_{\unlabeled,i}$.

On the other hand, querying a twin $\twin_{\unlabeled,i}$ has an associated computation cost that is the sum of: (a) cost of generating the twin-world and (b) cost of the generating propagation rays over locations on the road and all beams for creating the lookup table. Thus, the total cost for creating a twin is: 

\begin{equation}
\cost^{\twin_{\unlabeled,i}}_{Comp} = \cost^{\twin_{\unlabeled,i}}_{map} +  \cost^{\twin_{\unlabeled,i}}_{lookup},
\label{eqn:cost}
\end{equation}
where $\cost^{\twin_{\unlabeled,i}}_{map}$ is the cost to generate/import the maps for creating the world, and $\cost^{\twin_{\unlabeled,i}}_{lookup}$ is the cost of running ray tracing emulation to generate the lookup table, which is further broken down as $\cost^{\twin_{\unlabeled,i}}_{lookup} = \beam_{i} \times n_{\twin_{\unlabeled,i}}\times N_{j}^{m,i}  \times \cost_{prop}(\reflect_i) $.  
Here, $\cost_{prop} (\reflect_i)$ is the cost for calculating one propagation ray with $\reflect_i$ reflections that depends on the available compute resource. However, following the Wireless InSite manual~\cite{WI_references}, the cost for propagation path calculation depends on four emulation parameters as:
\begin{equation}
    \cost_{prop}(\reflect_i) \propto \frac{(\reflect_i+\transpath+1)!}{\transpath!\reflect_i!} \times w^{\diffract + 1},
\end{equation}
where $\reflect_i$ denotes the number of reflections for twin $\twin_{\unlabeled,i}$. Moreover, and $\transpath$ and $\diffract$ are the number of transmissions~(penetration) through the walls and diffractions, respectively. Finally, $w$ is the number of reflective surfaces~(walls for example) that are reached by the rays emitted from the Tx and completely depends on the map geometry.

\noindent\textbf{Creation of Lookup Table.}
\label{subsubsec:lookup_table}
After modeling all the propagation paths for different Rx location in each twin, we create a lookup table consisting of four parameters: (a) {\em Rx location index  $l_{j}\in L_{\twin_{\unlabeled,i}}$}, (b) {\em Twin index $\twin_{\unlabeled,i}$}, (c) {\em SNR values $\texttt{SNR}_{j}^{m,i}$ for all beams $t_m\in C^{Tx}_{i}$ of twin $\twin_{\unlabeled,i}$ over all locations $l_{j}\in L_{\twin_{\unlabeled,i}}$}, and (d) {\em computation cost 
$\cost^{\twin_{\unlabeled,i}}_{Comp}$ associated to twin $\twin_{\unlabeled,i}$}, hence represented as a dictionary: $LT (\twin_{\unlabeled,i}) = \{ (l_{j}, \twin_{\unlabeled,i},\texttt{SNR}_{j}^{m,i}, \cost^{\twin_{\unlabeled,i}}_{Comp})\}_{j=1}^{n_{\twin_{\unlabeled,i}}}$.

\vspace*{7pt}
\subsubsection{Real-time Multiverse Interaction}
\label{subsec:multiverse_interaction}
If it is determined by the vehicle that it encountered an unseen scenario $\env_\unlabeled$, it requests accessing to the lookup table $LT (\twin_{u,i})$ from the $\MV$. Recall that each of the twins $\{\twin_{\unlabeled,i}\}_{i=1}^{N}$ in the $\MV$ offers different levels of fidelity and computation cost. Moreover, the  communication constraints are vehicle-specific and depend on the data that needs to be delivered~(safety-critical tasks such as driving directives generation versus the media for example). To account for such constraints, we propose an algorithm to  jointly optimize the selection of twins from the $\MV$ and associated top-$K$ beam candidates.

\noindent{\bf Modeling Fidelity (probability of inclusion).} 
A simple way to model the fidelity is by benchmarking the performance of each twin with respect to the ground-truth from real world measurements. We define the probability of inclusion as the probability of the best beam from the ground-truth being in the top-K predictions from the $\MV$. Our observations indicate that the probability of inclusion for each twin is related to the sub-regions in the coverage area of the transmitter. Moreover, it is affected by the number of allowed beams~$K$. To that end, we leverage the empirical observations to model the probability of inclusion as $p(K, LT (\twin_{\unlabeled,i}),\env_\unlabeled,r)$, where parameters $K$, $LT (\twin_{u,i})$, $\env_\unlabeled$, and $r$ denote that number of selected beams, set of lookup table entries for twin $\twin_{\unlabeled,i}$, current unseen scenario $\env_\unlabeled$, and the region in which the receiver is located, respectively.

\noindent{\bf Communication and Computation Costs.} 
We use the state-of-the-art 5G-NR standard~\cite{barati2020energy} to model the communication cost for sweeping the top-$K$ beams as:
\begin{equation}
    \cost^{\texttt{5G-NR}}(K) = T_p \times \left \lfloor\frac{ K - 1}{32} \right \rfloor + T_{ssb},
    \label{eq:T_nr}
\end{equation}
Here, $T_p=20\,ms$ and $T_{ssb}=5\,ms$ correspond to the periodicity and synchronization signals~(SS) burst duration that are used for sweeping $K$ beams, respectively. On the other hand, the computation time is a function of computation power for a given twin within the $\MV$. The computation cost $\cost_{Comp}^{\twin_{u,i}}$ for $\twin_{u,i}$ is retrieved by querying the lookup table $LT (\twin_{u,i})$, which is derived from Eq.~\ref{eqn:cost}.

\noindent{\bf Optimization.}
We solve the following optimization problem to determine the optimum twin from the $\MV$ and corresponding $K$ beams as:
\begin{subequations}
\label{eq:opt_twin_selection}
\begin{align}
    \underset{K,\twin_{u,i}}{\operatorname{max}} &~~ p(K,
    LT (\twin_{u,i})
    ,\env_\unlabeled,r)+\frac{\alpha}{2} (1-\frac{\scriptstyle\cost^{\texttt{5G-NR}}\scriptstyle( K)}{\scriptstyle\cost_{Comm}})(1-\frac{\scriptstyle\cost_{Comp}^{\twin_{u,i}}}{\scriptstyle\cost_{Comp}}),\\
    \text{s.t.} &\quad \cost^{\texttt{5G-NR}}(K) < \cost_{Comm},\\
                &\quad \cost_{Comp}^{\twin_{u,i}} < \cost_{Comp},\\
                &\quad r^b\leq r \leq r^e,\\
                &\quad \env_\unlabeled \in \{\env_\unlabeled\}_{u=1}^{\numberunseen},\\
                &\quad \alpha>0.
\end{align}
\end{subequations}

In Eq.~\ref{eq:opt_twin_selection}, the first term in objective forces the optimization problem to choose twins with higher fidelity (probability of inclusion). As opposed to this, the second term prevents the selection of unnecessarily high $K$ values and more complex twins for time sensitive applications. The control parameter $\alpha$ in Eq.~\ref{eq:opt_twin_selection} weights the importance between these two terms in the objective function. Moreover, the parameters $\cost_{Comm}$ and $\cost_{Comp}$ are user defined communication constraints at the vehicle and computation constraint at the edge. The computational complexity for running this optimization is in the order of $O(N)$ with $N$ denoting the number of twins. After solving the Eq.~\ref{eq:opt_twin_selection}, the optimum twin and the associated top-$K$ beams are determined. The vehicle then sweeps through those suggested top-$K$ beams in the real world, obtains the optimum beam, and starts the transmission.


\subsection{Fine-Tuning with Multiverse Ground-Truth}
\label{subsec:realworld_augmentation}
When the vehicle uses its local DL models for beam selection, the local multimodal sensor data is used for inference~(see Sec.~\ref{sec:ml_beam_selection}). This results in near-real time prediction of the beam, due to locality of all actions. However, pure machine learning-based methods suffer from the adaptation to unseen environments, which undermines the accuracy of the model prediction. We propose to incorporate the emulation output from the $\MV$ to fine-tune the local DL model at the vehicle. Once a twin within the $\MV$ is invoked to obtain the optimal beam, the ray tracing based emulation outputs become the ground-truth labels and are communicated back to the real world. We propose to pair these labels with the local sensor data at the vehicle to fine-tune the local model, periodically, few epochs of training for example. As a result, the local model can account for the previously unseen environment in the future, instead of continuous triggering of twins from the $\MV$. Following the notations of Sec.~\ref{sec:ml_beam_selection}, the fine-tuned  model parameters ${\hat{\theta}}$ is obtained by minimizing the loss $\loss(\hat{\theta}; \env_\unlabeled)$ on the unseen scenario $\env_\unlabeled$, where  $\loss(\hat{\theta}; \env_\unlabeled) := \frac{1}{n_\unlabeled}\sum_{j=1}^{n_\unlabeled} [\ell(f_{\hat{\theta}}({X}_{\unlabeled,j}^{\ldr}, {X}_{\unlabeled,j}^{\img}, {X}_{\unlabeled,j}^{\crd}), Y_{\unlabeled,j})]$. Here, $Y_{\unlabeled,j}$ is the labels derived from twin $\twin_{\unlabeled,i}$ for unseen scenario $\env_\unlabeled$ ($|Y_{\unlabeled,j}|=n_\unlabeled$) and ${X}_{\unlabeled,j}^{\ldr}, {X}_{\unlabeled,j}^{\img}, {X}_{\unlabeled,j}^{\crd}$ are the real sensor data recorded at the vehicle.

\subsection{End-to-End System}
Given a test sample, we first identify if it is either in-distribution or out-of-distribution, corresponding to seen and unseen scenarios, respectively. There are state-of-the-art works where kernel density-based~\cite{subramaniam2006online}, nearest neighbour-based~\cite{angiulli2002fast,zhang2006detecting}, or reconstruction-based~\cite{markou2003novelty} methods are proposed for identifying unseen scenarios. For example, one of our previous works~\cite{Zifeng_2019} focuses on using statistical metrics recorded on the training data to detect unreliable predictions. This includes metrics that characterize the prediction probability confidence and ratio of correctly classified samples for each class, beams in our case. These statistics are generated at the training phase and are used as a threshold at inference to identify the outliers. We denote the outlier detection operation by $\gamma (.)$.

Consider a test sample with $\mathbb{X}_{m}^{\ldr}, \mathbb{X}_{m}^{\img}, \mathbb{X}_{m}^{\crd}$ as the LiDAR, image and coordinate samples, respectively. If the sample is detected as in-distribution, we use the DL-based method to locally predict the optimum beam using the sensor data. If the sample is out-of-distribution, the vehicle submits a request to get access to the $\MV$ of twins and solves Eq.~\ref{eq:opt_twin_selection} to select the best twin and top-$K$ beams. Formally, 
\begin{equation}\label{eqn:p1}
    \begin{split}
    t^{*} &= 
    \begin{cases}
       \sigma(f_{\theta}({\mathbb{X}}_{\labeled,m}^{\ldr}, {\mathbb{X}}_{\labeled,m}^{\img}, {\mathbb{X}}_{\labeled,m}^{\crd})) & \text{if}~ \gamma ({\mathbb{X}}_{\labeled,m}^{\{\ldr,\img,\crd\}})=D_{\text{ID}} \\
       \text{Solve Eq.~\ref{eq:opt_twin_selection}} & \text{if}~\gamma ({\mathbb{X}}_{\labeled,m}^{\{\ldr,\img,\crd\}}) = D_{\text{OOD}}\\
     \end{cases}
    \end{split}
\end{equation}
where $\sigma$ denotes {\em softmax} activation, $D_{\text{ID}}$ implies that the data sample belongs to the seen scenarios (in-distribution) and $D_{\text{OOD}}$ denotes that $\gamma (.)$ has identified that the sample is out-of-distribution (OOD), i.e. unseen scenario. For out-of-distribution samples, our framework revolves around (i) creating a $\MV$ in the edge server of base station, and (ii) interacting between the $\MV$ and real world~(see Fig.~\ref{fig:system_model}). The {\em trigger $\MV$} module calls for the $\MV$ $\multiverse$ from the edge, following Eq.~\ref{eqn:p1}. The edge has a few in-library twins within the $\MV$. This includes the twins that are encountered before and are already available at the edge. If the requested twin from the vehicle is already available at the edge, a look up table with information about the locations of the receiver and
SNR of all beams are sent in the downlink through a multi-Mbps sub-6~GHz wireless control channel. The vehicle then exploits this look up table to jointly optimize the selection of twin and associated top-$K$ beam candidates using Eq.~\ref{eq:opt_twin_selection}. If the requested twin is not available, the sensor data of the vehicle is sent in uplink to the edge, where the ray tracing tool runs to generate a look up table for the unseen environment. The output look up table is added to twin family in the $\MV$. Note that if the requested twin is already available at the edge, the $\MV$ provides an output in competitive time compared to exhaustive search. If not, the computationally extensive ray tracing must be run at edge which might not provide an output in short contact times in vehicular network. However, the simulation output can be later used to account for the new vehicles, as they encounter the same scenario.

\section{Experiment Design}
\label{sec:multiverse_creation}
We consider the real world to be a V2I setting in an urban canyon of a metro city. We deploy the $\MV$ using a ray tracing simulation tool, Wireless InSite~($\WI$) by RemCom~\cite{WI_webpage}. $\WI$ is capable of modeling complex electromagnetic propagation patterns and communication channels in 3-dimensional~(3D) urban, indoor, rural, and mixed path environments.
\begin{figure*}[t]
     \centering
    \begin{subfigure}[b]{0.245\textwidth}
         \centering
         \includegraphics[width=\textwidth]{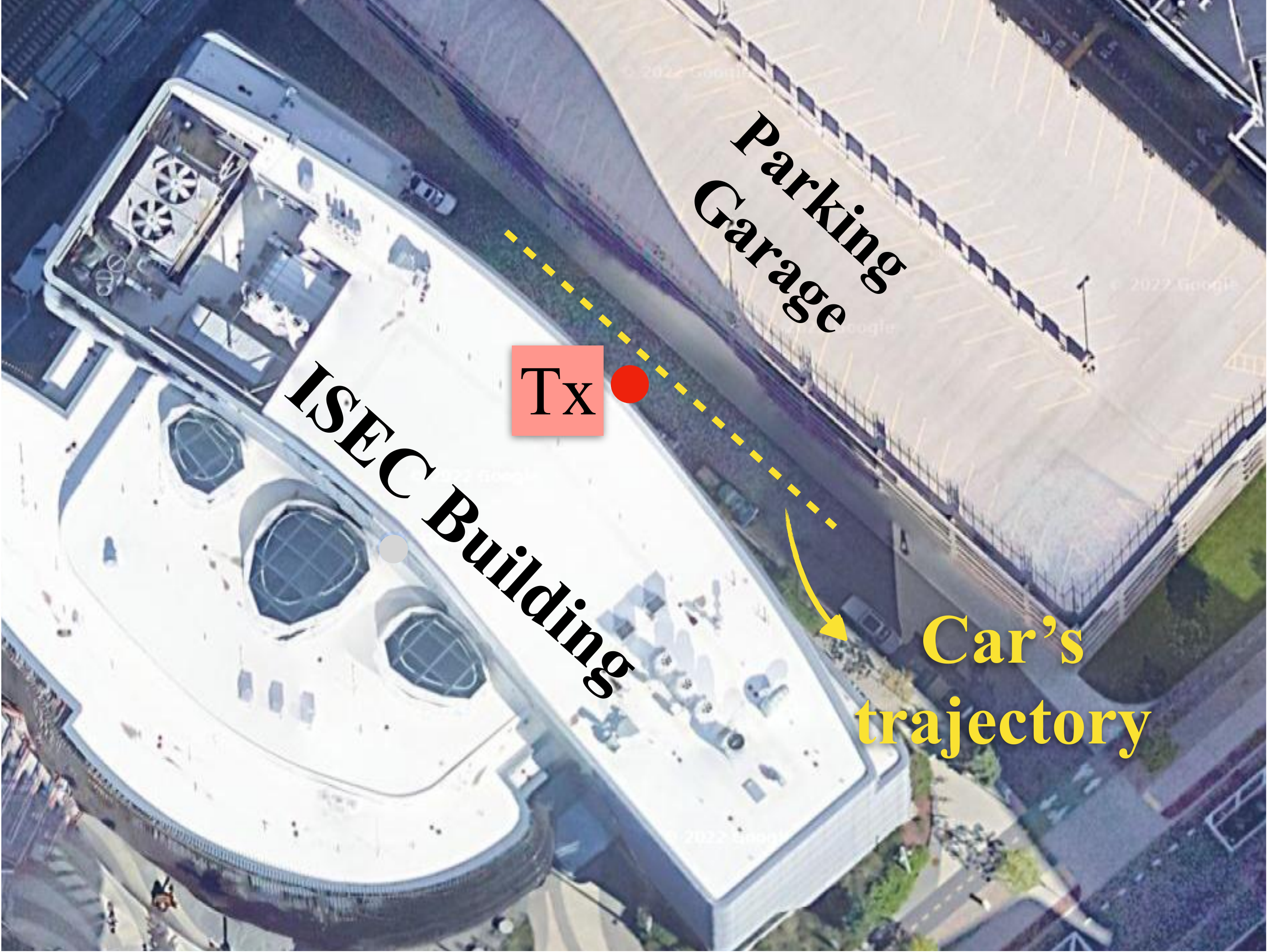}
         \caption{}
         \label{fig:exp_1}
     \end{subfigure}
     \begin{subfigure}[b]{0.245\textwidth}
         \centering
         \includegraphics[width=\textwidth]{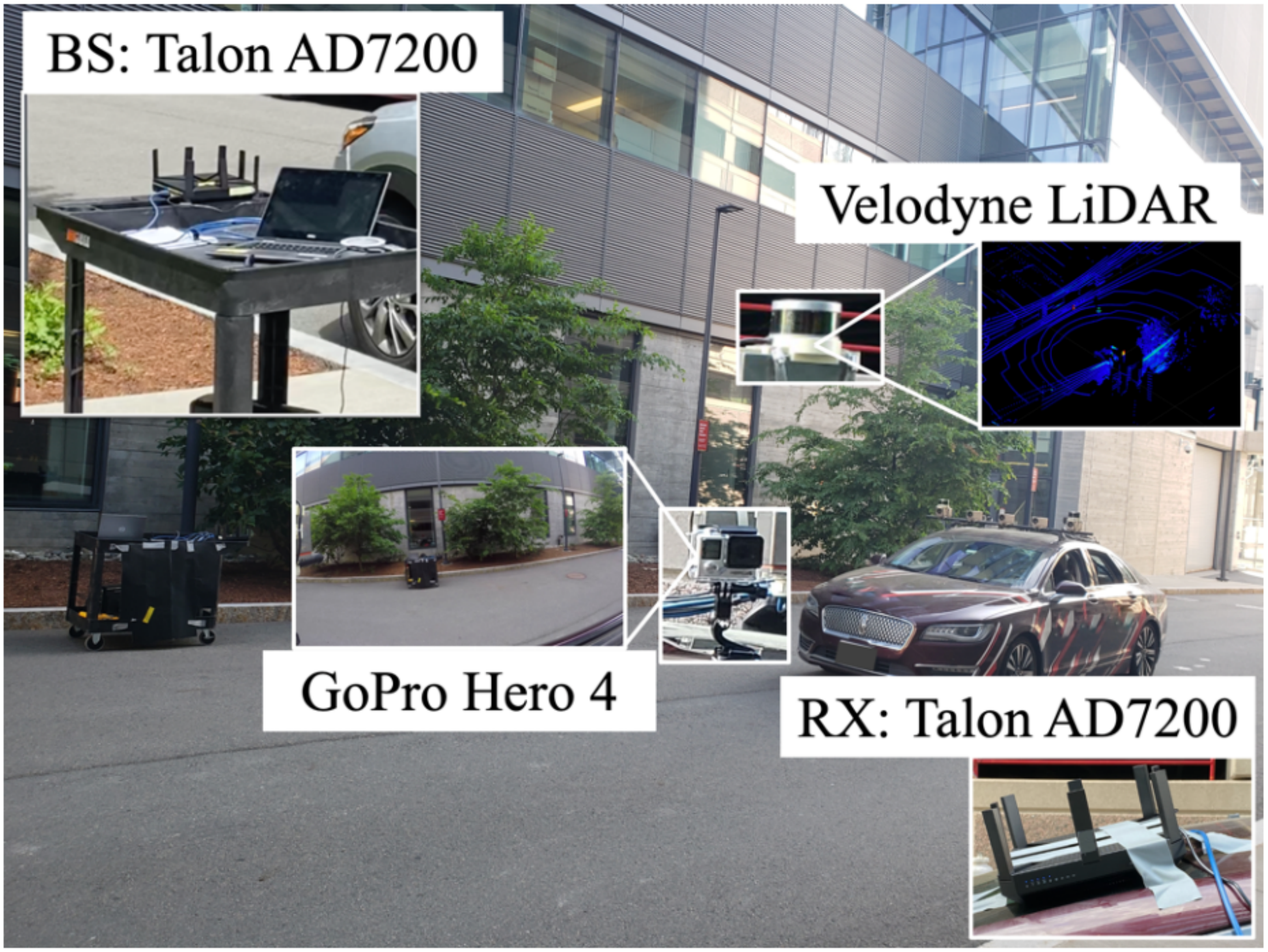}
         \caption{}
         \label{fig:exp_2}
     \end{subfigure}
     \begin{subfigure}[b]{0.245\textwidth}
         \centering
         \includegraphics[width=\textwidth]{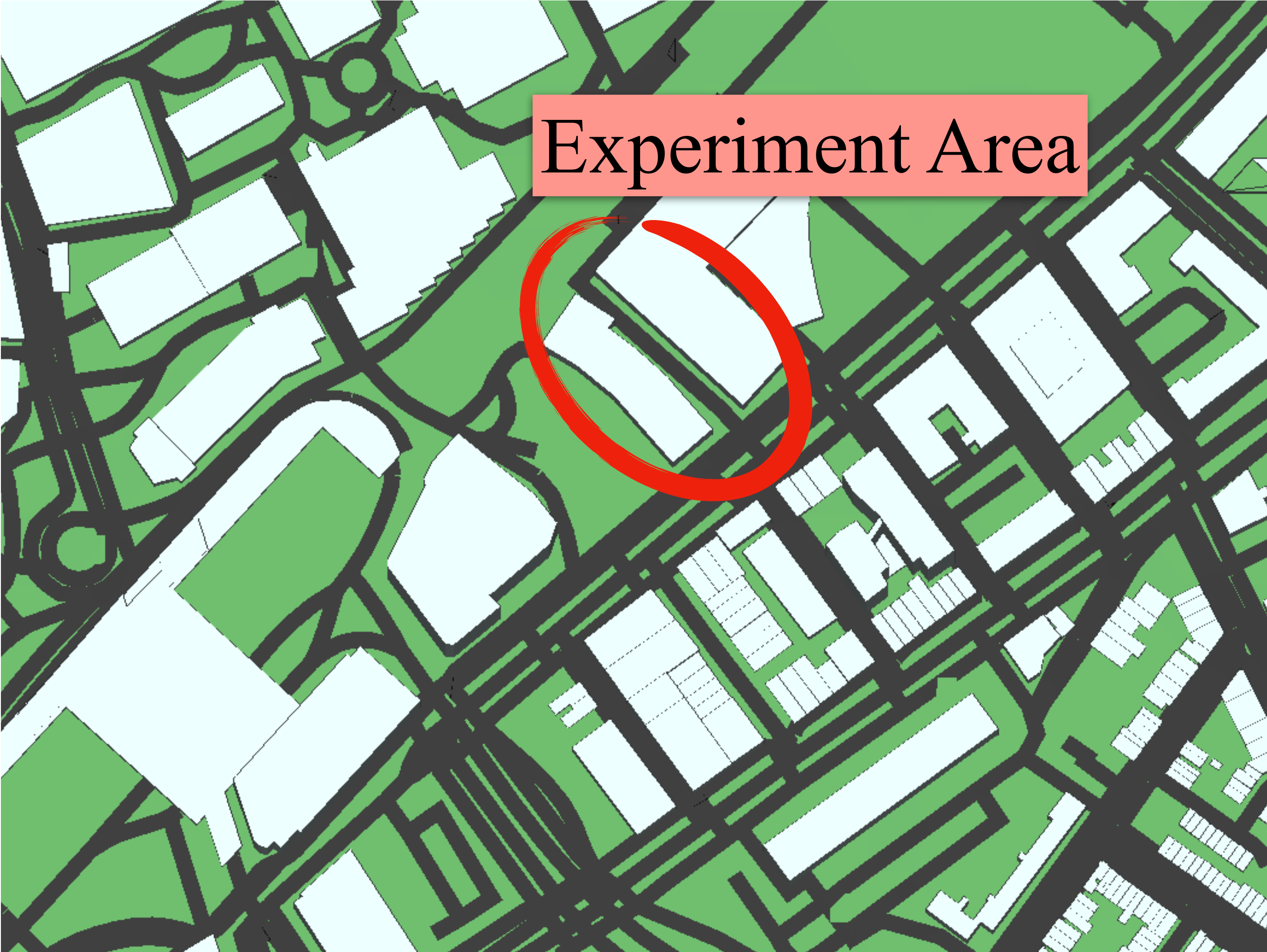}
         \caption{}
         \label{fig:exp_3}
     \end{subfigure}
      \begin{subfigure}[b]{0.245\textwidth}
         \centering
         \includegraphics[width=\textwidth]{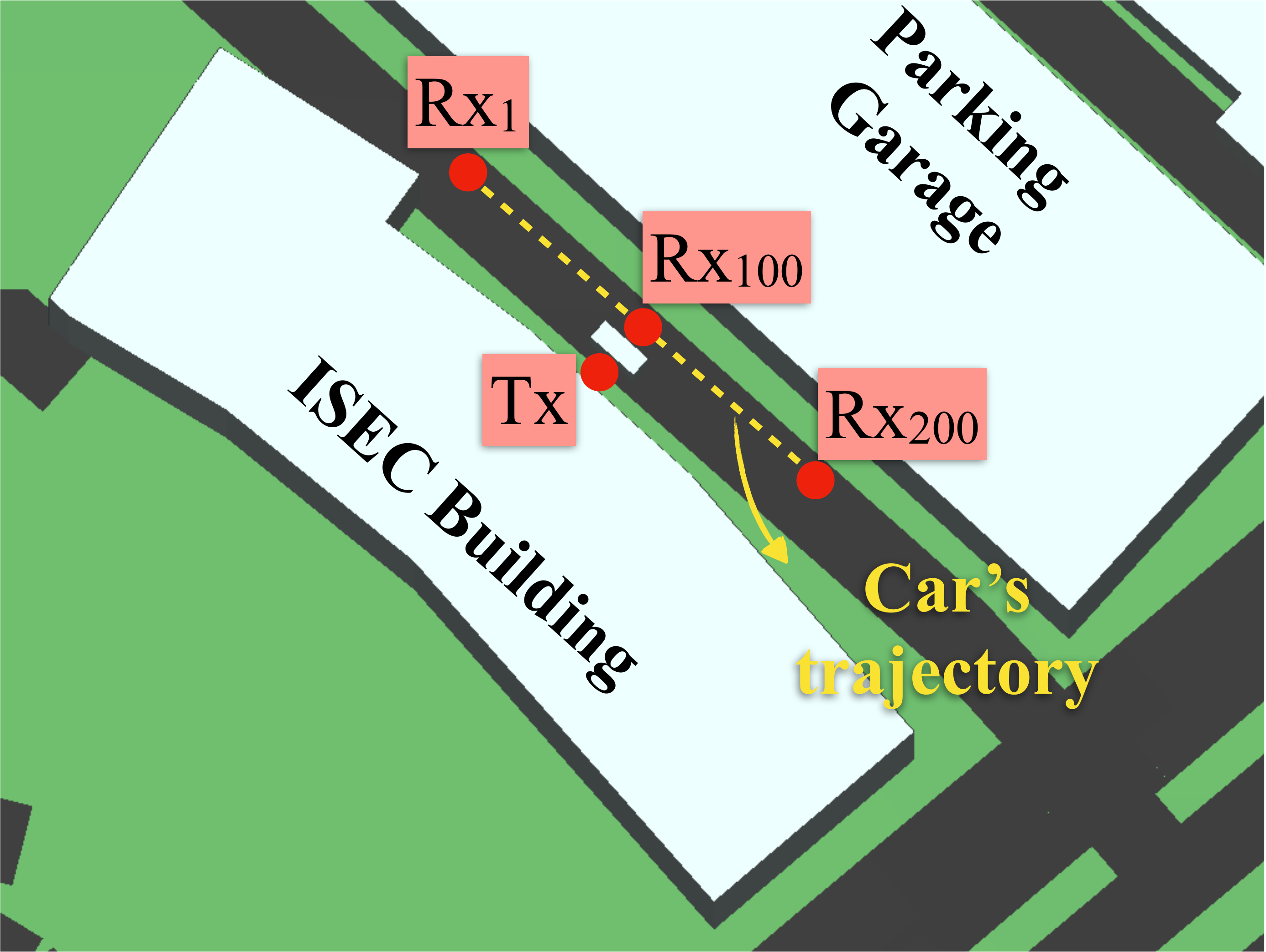}
         \caption{}
         \label{fig:exp_4}
     \end{subfigure}
     \caption{
     (a) Experiment location in real world; (b) FLASH experiment setup~\cite{salehi2022flash}; (c) Experiment area in the virtual world (\emph{twin}), showing the map beyond the area of interest in downtown Boston; (d) Simulation location in \emph{twin}, showing the first, middle, and last sample points, in order. The 3D box between Tx and Rx\textsubscript{100} models the car obstacle in the NLOS scenario.}
     \label{fig:world_comparison}
\end{figure*}

\subsection{Real world: Creation of Ground-Truth}
\label{sec:dataset_flash} 
We use a publicly available mmWave beam selection dataset called FLASH~\cite{flash_dataset} to generate the ground-truth from the real world. This dataset is captured within downtown Boston on an asphalt road flanked with high rising buildings~(see Fig.~\ref{fig:exp_1}). The dataset includes synchronized sensor data from on-board GPS, a GoPro Hero4 camera, and a Velodyne LiDAR, all mounted on a 2017 Lincoln MKZ Hybrid autonomous car~(see Fig.~\ref{fig:exp_2}). The ground-truth includes the received signal strength indicator~(RSSI), which is acquired using the Talon AD7200 mmWave radio~\cite{Steinmetzer_2017}, with a pre-defined codebook of 34 beams. The dataset includes both LOS and NLOS scenarios with approximately 5K and 3K samples each, respectively.
The LOS scenario corresponds to the vehicle~(i.e. receiver) passing in front of the transmitter without any obstacle, while in NLOS scenario, a vehicle is located in front of the Tx, blocking LOS.

\begin{figure}
\centering
  \includegraphics[width=\linewidth]{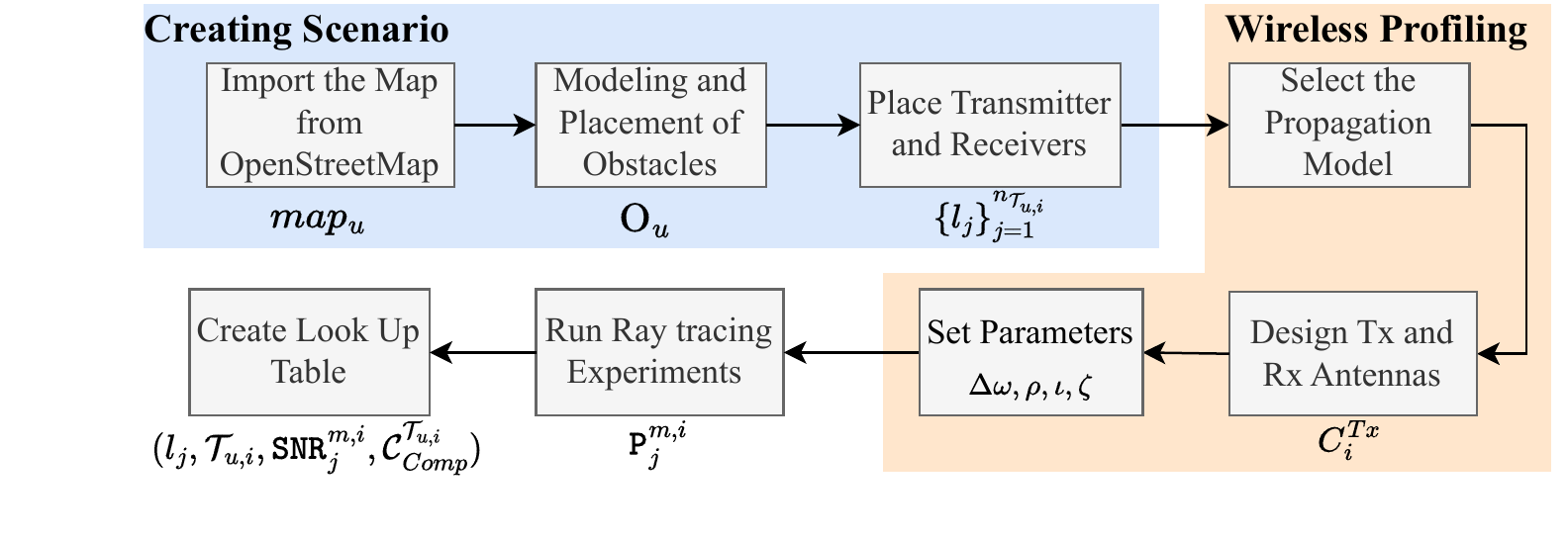}
  \caption {A typical workflow in Wireless InSite (WI) for creating a twin (suppose $\twin_{u,i}$ for scenario $\env_u$) in the $\MV$.}
  \label{fig:wi_pipeline}
\end{figure}

\subsection{Multiverse Setup}
In this section, we describe our approach for creating the scenarios and setting the RF parameters. The modular steps to create different twins is depicted in Fig.~\ref{fig:wi_pipeline}.

\subsubsection{Creating Scenarios in Wireless InSite} The Wireless InSite ray tracing tool allows users to create a digital environment with 3D maps. We replicate the testbed of FLASH dataset in WI, using the closest materials and positioning for the objects. A sample map visual for the digital world is provided in Fig.~\ref{fig:exp_3}. 

\noindent{\bf Import the Map.} 
In $\WI$, a digital world can have various features, which are stored as layers, e.g. buildings, roads, water bodies, and foliage. Buildings are modeled as boxes that can have different material features, e.g. concrete, glass, metal, providing near-realistic ray reflection and penetration. The first step in creating a scenario is to obtain an appropriate map.  OpenStreetMaps~\cite{OpenStreetMap_webpage} provides such data, which we process in Blender~\cite{Blender_website} as geodata for $\WI$. We confirm that the road and valley width as well as building dimensions in the digital world match with the ones in the real world.

\noindent{\bf Modeling and Placement of Obstacles.} In WI, the dimension and coordinate of obstacles can be set, as desired. While a LOS scenario is ready for simulation runs after setting up the map and adding all the aforementioned details, there is one more step to finalize a NLOS scenario, modeling the obstacles. Following the FLASH testbed, where the obstacle is a 2018 Toyota Camry with the dimensions of $1.44m\times1.84m\times4.88m$, height, width, and length, respectively, we model the obstacle in $\WI$ as a metal box with the same dimensions and place it $35cm$ away from the Tx, shown as white box in Figs.~\ref{fig:exp_4}.

\noindent{\bf Placement of Tx and Rx.} Following the FLASH testbed~\cite{salehi2022flash}, the Tx antenna is placed on a cart with the height of $75cm$. Moreover, the height of Tx itself is $20cm$. Thus, we set the Tx height to be $95cm$ in $\WI$. The receiver car in the FLASH dataset has a height of $147.5cm$. Thus, we set the Rx height to be $167.5cm$ in the simulation environment. In each simulation, we have one Tx location and one or more Rx locations (denoted as $L_{\twin_{\unlabeled,i}}$ for $\twin_{\unlabeled,i}$). We consider two set of scenarios, including LOS~($\env_1$) and NLOS~($\env_2$)

\subsubsection{Setting RF Parameters in Wireless InSite} ~\\
\noindent{\bf Selection of Propagation Model.} $\WI$ provides a number of propagation models. We use \emph{X3D} because it provides high fidelity simulations, taking the following channel parameters into account: 3D ray tracing, indoor-outdoor suitability, atmospheric attenuation (temperature, pressure, humidity), reflection angle dependent reflection coefficients, and MIMO system analysis. Computation-wise \emph{X3D} uses GPU acceleration.

\noindent {\bf Designing the Tx and Rx Antennas.} In the real world experiments~\cite{flash_dataset}, the Talon AD7200 routers are used with Qualcomm’s QCA9500 FullMAC IEEE 802.11ad Wi-Fi chip for both Tx and Rx. From the open source characterization of the antenna pattern~\cite{steinmetzer2017compressive}, we retrieve the SNR measurements in 3D for 802.11ad standard~(60GHz band) to create the Tx and Rx antenna patterns in $\WI$. The 3D legacy azimuth ($\phi$) and elevation ($\theta$) measurements span $[-90^{\circ}, 90^{\circ}]$ and $[0^{\circ}, 32.4^{\circ}]$, respectively, having 101 and 10 sample points. We show the virtual representation of $\phi$ and $\theta$ in Fig.~\ref{fig:Ant_sim_b}. Moreover, we match the antenna orientations of WI and the ones in the FLASH dataset on the xy-plane, having the azimuth ($\phi = 0^{\circ}$) axis of Tx and Rx parallel~(See Fig.~\ref{fig:twinsALL_LOS_NLOS_rays}).

\noindent{\bf Selection of $\Delta\omega$, $\reflect$, $\transpath$, and $\diffract$}. 
$\WI$ allows users to select ray spacing $\Delta\omega$ (angle between two adjacent emitted rays), maximum number of reflections $\reflect$, maximum number of transmissions $\transpath$, and maximum number of diffractions $\diffract$. In our experiments, we select ray spacing to be $\Delta \omega = 0.25^{\circ}$. We set $\reflect$ to be either 1 or 3, because we observe that given the geometry of the environment and the effect of reflections on the received power, this number of reflections is enough to deliver signals to the receivers. We choose the $\transpath$ to be 0, because the data collection environment (urban canyon) is mostly concrete on the sides and the floor is asphalt, which typically do not allow mmWave signals to penetrate through. Finally, $\diffract$ is set to be 1, because after 2 diffractions, mmWave signal strength is negligible.

\subsubsection{\bf Ray Tracing in Wireless InSite.} For each scenario, we perform ray tracing over Rx locations and 34 Tx beams for each twin. In the experiments, the angle between two rays is set as $\Delta \omega = 0.25^{\circ}$ and rays are transmitted at all directions in the Euclidean space, following the antenna pattern specifications. Since the rays are emitted in discrete angles and spread as they travel through space, it is not guaranteed that rays would be received at exact Rx location. Thus, in order to ensure that received rays are correctly estimated, $\WI$ puts a sphere around the receiver point, with the radius \textbf{r}. Depending on how large that \textbf{r} is, several rays might pass through the sphere. There are two steps to identify which ray to select: i) rays are sorted according to the geometry faces they interact with and the similar ones are eliminated in order to prevent over-predicting the delivered energy at the receiver and ii) the closest ray to the actual Rx point is selected. According to WI manual, $\boldsymbol{r}=\Delta \omega \times D_{max}$, where $D_{max}$ is the coverage range of Tx~($\approx 20m$ in FLASH). Thus, we estimate $\boldsymbol{r} \approx 8.73cm$. The X3D model then applies Exact Path Calculations~(EPC), which adjusts the reflection and diffraction points within the sphere so that the selected rays pass through the exact receiver points. We consider the highest received power value of the received ray from each Tx beam at each of the $L_{\twin_{u,i}}$ Rx locations for twin $\twin_{u,i}$.


\subsection{Multiverse Setup: Twins in Multiverse}
We create the $\MV$ with three twins each having different cost and precision. All twins have the same materials for the surrounding buildings and foliage, antenna orientations, transmitted waveforms, transmission power, Tx and Rx antenna patterns, coordinates for the Rx sample points, noise power, and antenna sensitivity. To create differences between \emph{twins}, we change (i) environment used in the twin world, (ii) number of allowed reflections~($\reflect$) in simulations, and (iii) number of Rx locations, $L_{\twin_{u,i}}$. The common parameters in the digital world creation are given in Tab.~\ref{tab:VWorldParam}, whereas the differences between twins are highlighted in Tab.~\ref{tab:TwinDiff}.

\begin{table}[t!]
\centering
\begin{tabular}{|l||l|}
\hline
Materials & \begin{tabular}[c]{@{}l@{}}Buildings: Concrete\\ Road: Asphalt\\ Foliage: Grass\end{tabular}\\ \hline
Antenna orientation & T\textsubscript{x} and R\textsubscript{x} facing opposite directions\\ \hline
Waveform & $f_{c} = 60 GHz$, BW = $2.16GHz$\\\hline
Tx power& $24 dBm$~\cite{TalonSpec_website}\\ \hline
Tx and Rx pattern source  & Talon antenna measurements~\cite{AntennaPatterns_website}\\ \hline
Tx height& On a cart, $0.95 m$\\ \hline
Rx height& On a car, $1.645 m$\\ \hline
Noise power~($\mathcal{N}$)& $-100.99 dBm$\\ \hline
Antenna sensitivity & $-250dBm$\\ \hline
Ray spacing & $\Delta \omega = 0.25^{\circ}$\\ \hline
$\diffract$ & 1 \\ \hline
$\transpath$ & 0 \\ \hline
Scenarios & $\env_1$ \&  $\env_2$ \\\hline
\end{tabular}
  \caption {Multiverse parameters.}
  \label{tab:VWorldParam}
\end{table}

\begin{table}[t!]
\centering
\begin{tabular}{|c|c|c|c|}
\hline
 Twins & $\first$ ($\twin_{u,1}$)   & $\second$ ($\twin_{u,2}$)   & $\third$ ($\twin_{u,3}$) \\\hline\hline
Environment & Open space & Urban canyon & Urban canyon \\ \hline
Imported Map & Random & Boston & Boston \\ \hline
$\reflect_i$ & 3  & 1  & 3  \\ \hline
$\cost^{\twin_{u,i}}_{lookup}$ (hour) & $\sim0.4$  & $\sim50$  & $>100$  \\\hline
\#Rx Locations& $n_{{\twin}_{u,1}}=1$   &$n_{{\twin}_{u,2}}=200$   &$n_{{\twin}_{u,3}}=200$  \\\hline
\end{tabular}
  \caption {Differences between twins.} 
  \label{tab:TwinDiff}
\end{table}

\subsubsection{$\first$ Twin $\twin_{u,1}$}
\label{subsec:Twin1} 
Creating a precise map is expensive and may not be available all the time. Thus, in order to show the gain of precise geometry~(which we create for the next twins) and to form a baseline for our evaluations, we use an \emph{open area} environment as the \first~twin~(excluding the buildings present in the FLASH testbed). The \emph{Open area} environment is not a free-space, as reflections from the far surrounding buildings in a city is still allowed. We set the Tx and only one Rx around center region of the \emph{open area}, locating them $4.33m$ away, corresponding to the minimum Tx-Rx distance in the FLASH testbed, and record the SNR for 34 beams. We use only one Rx location, because in the \emph{open area} experiments, different Rx locations are negligibly affected by the surroundings, due to high attenuation. Such an experiment setting significantly cuts both time and computation in ray tracing and forms a basis for our next \emph{twin} evaluations.
For this twin, we consider three reflections in our simulations. We design \first~twin for both the LOS ($\env_1$) and NLOS ($\env_2$) scenarios, with $\beam_1 = 34$. For the $\env_2$, we place the obstacle between Tx and Rx, $35cm$ away from the the Tx, parallel to the situation in the FLASH experiments. To account for different regions on the road, we first compute the angle between the Tx and Rx locations in FLASH. We then multiply the output from open area experiment with the antenna patterns from Talon router at the identified angle, for each beam.

\noindent{\bf Specific characteristics.} Following the notations of Sec.~\ref{sec:proposed_method}, the \first~twin~($\twin_{u,1}$) has: (a) $map_1$ = \emph{open area}, (b) $\mathrm{O}_1 $ = retrieved from FLASH~\cite{flash_dataset}, (c) $\reflect_1$ = $3$, and (d) $|L_{\twin_{u,1}}|$ = $1$.

\subsubsection{$\second$ Twin $\twin_{u,2}$}
\label{subsec:Twin2} 
For designing this twin, we switch the simulation environment to the location where the FLASH experiments took place. Creating the realistic environment and using antenna patterns from real life measurements brings a trade-off between significantly increased computation time and precision in beam selection and received power.
We set the Tx point in $\WI$ by precisely measuring the Tx location in the FLASH experiment location. Moreover, we import the buildings to $\WI$ with the same geometry as of the real world. In $\WI$ simulations, we collect received beam power at 200 consecutive points, which are uniformly distributed over a 40 meter trajectory, Tx being $4.33m$ away from the trajectory's middle point (see Fig~\ref{fig:exp_4}). We allow one reflection in the simulations to save on computation time and realize the effects of simpler settings that mostly allow LOS communication. For \second~twin, we have both the the LOS~($\env_1$) and NLOS~($\env_2$) scenarios, with $\beam_2 = 34$. For $\env_2$, we again place the obstacle between Tx and Rx, $35cm$ away from the the Tx, parallel to the situation in the FLASH experiments.

\noindent{\bf Specific characteristics.} The 1-reflection twin~($\twin_{u,2}$) has: (a) $map_2$ = \emph{Boston}, (b) $\mathrm{O}_2 $ = retrieved from FLASH~\cite{flash_dataset}, (c) $\reflect_2 = 1$, and (d) $|L_{\twin_{u,2}}|$ = $200$.

\subsubsection{$\third$ Twin $\twin_{u,3}$}
\label{subsec:Twin3} 
This twin creates a more comprehensive setting in the digital world in order to provide a more precise beam profiling, even in NLOS cases. We consider three reflections in the simulations to cope with potentially dense reflective environments. Like $\twin_{u,2}$ in Sec.~\ref{subsec:Twin2}, we run the simulations at the same place as the FLASH dataset was collected. We use 1 Tx, 200 Rx locations, and the same obstacle settings as in $\twin_{u,2}$ for both the $\env_1$ and $\env_2$ scenarios.

\noindent{\bf Specific characteristics.} Finally, for 3-reflection twin~($\twin_{u,3}$): (a) $map_3$ = \emph{Boston}, (b) $\mathrm{O}_3 $ = retrieved from FLASH~\cite{flash_dataset}, (c) $\reflect_3$ = $3$, and (d) $|L_{\twin_{u,3}}|$ = $200$.

\begin{figure}
     \centering
    \begin{subfigure}[b]{0.24\textwidth}
         \centering
         \includegraphics[width=\textwidth]{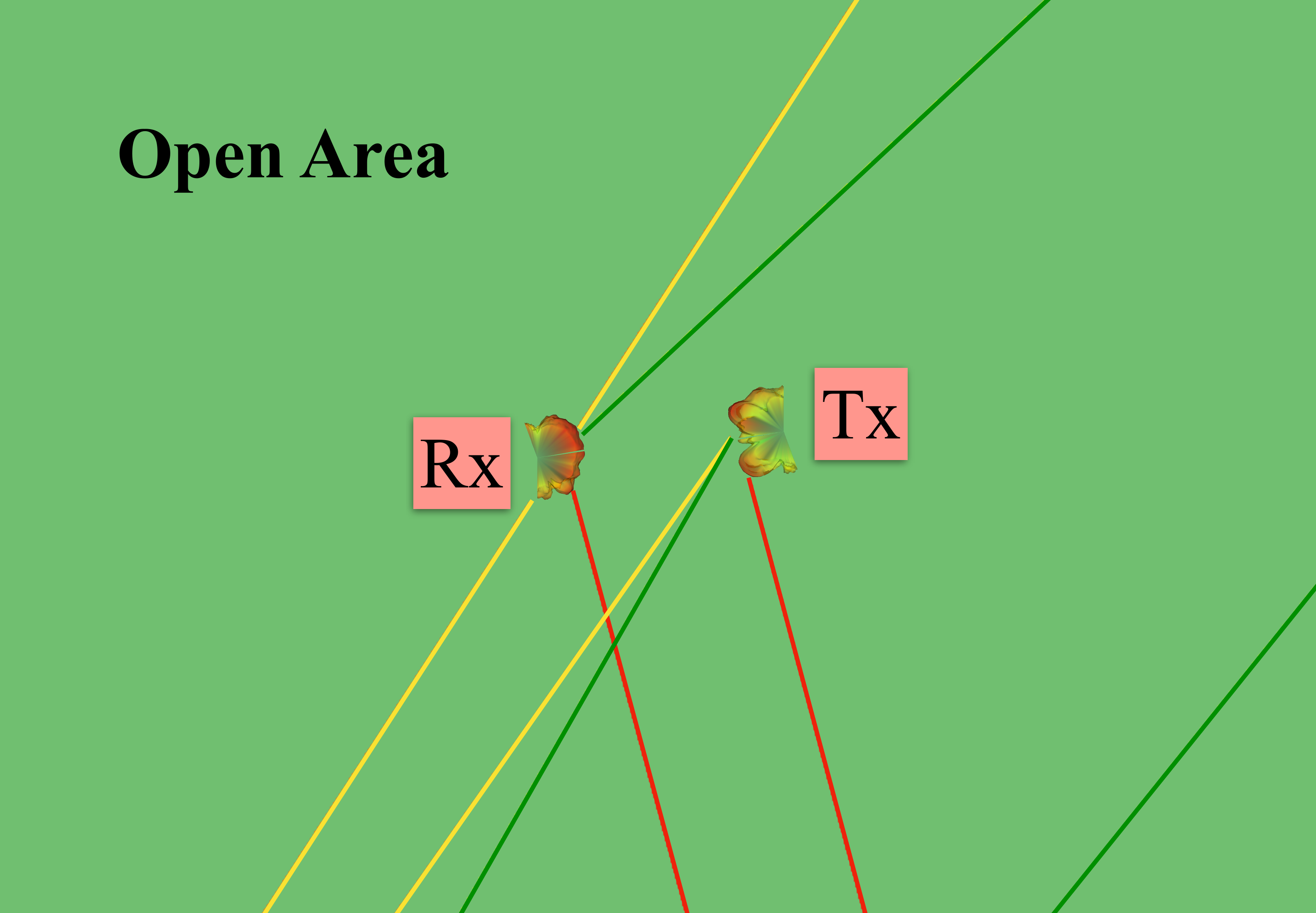}
         \caption{$\twin_{u,1}$ rays for $\env_1$.}
         \label{fig:twin1_LOS_rays}
     \end{subfigure}
     \vspace{10pt}
     \begin{subfigure}[b]{0.24\textwidth}
         \centering
         \includegraphics[width=\textwidth]{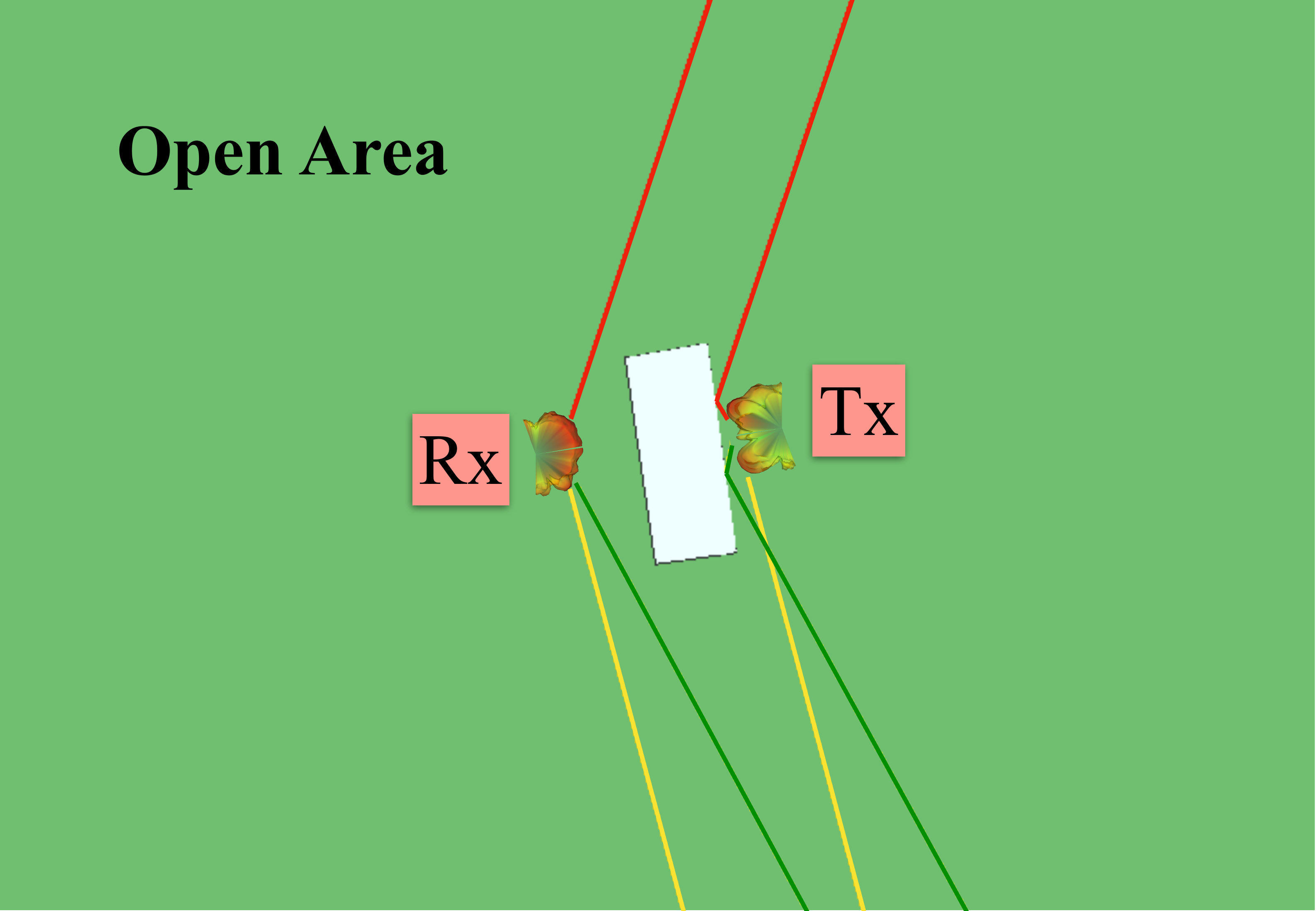}
         \caption{$\twin_{u,1}$ rays for $\env_2$.}
         \label{fig:twin1_NLOS_rays}
     \end{subfigure}
     \vspace{10pt}
     \begin{subfigure}[b]{0.24\textwidth}
         \centering
         \includegraphics[width=\textwidth]{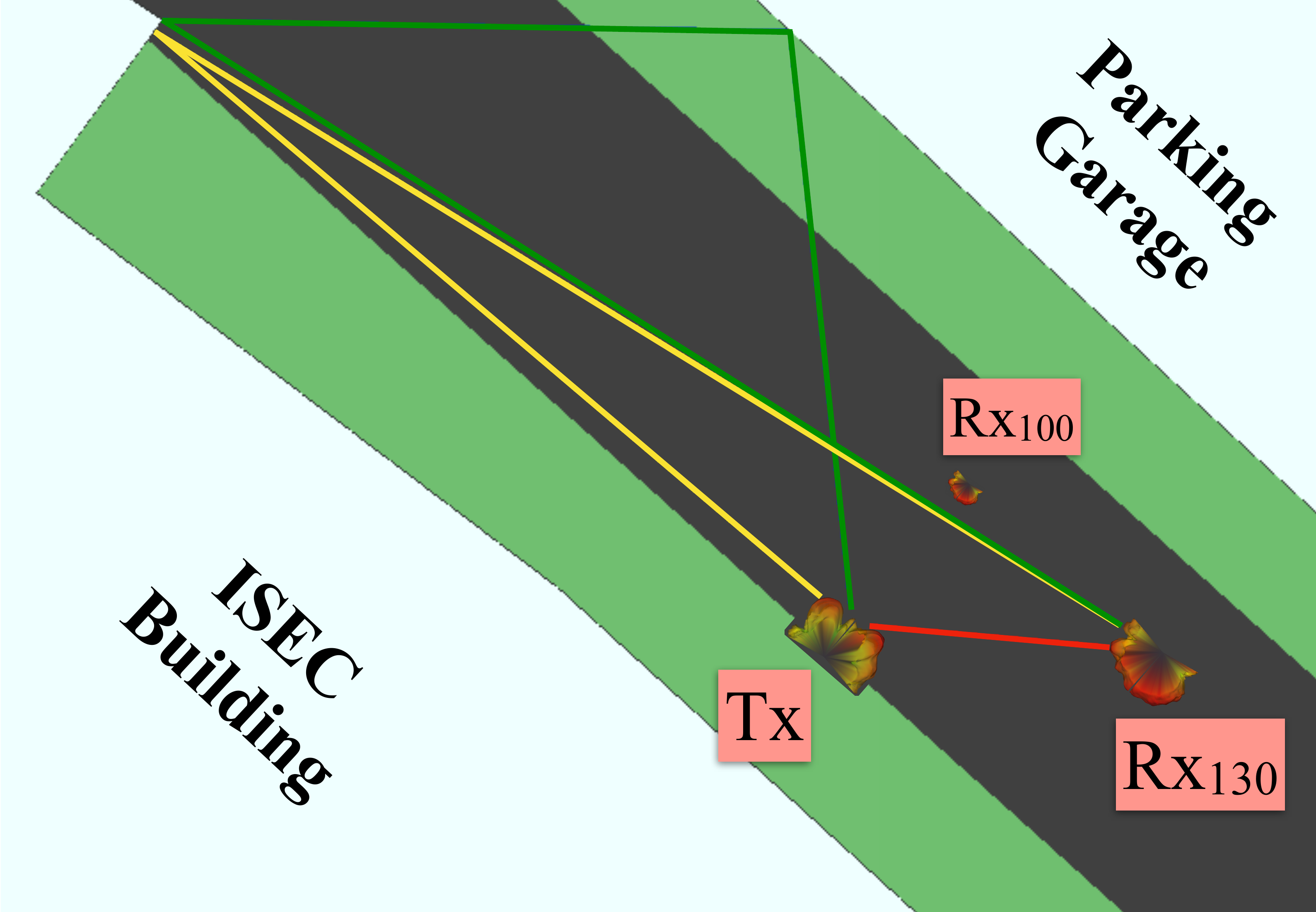}
         \caption{$\twin_{u,2}$ rays for $\env_1$ at Rx$_{130}$.}
         \label{fig:twin2_LOS_rays}
     \end{subfigure}
     \begin{subfigure}[b]{0.24\textwidth}
         \centering
         \includegraphics[width=\textwidth]{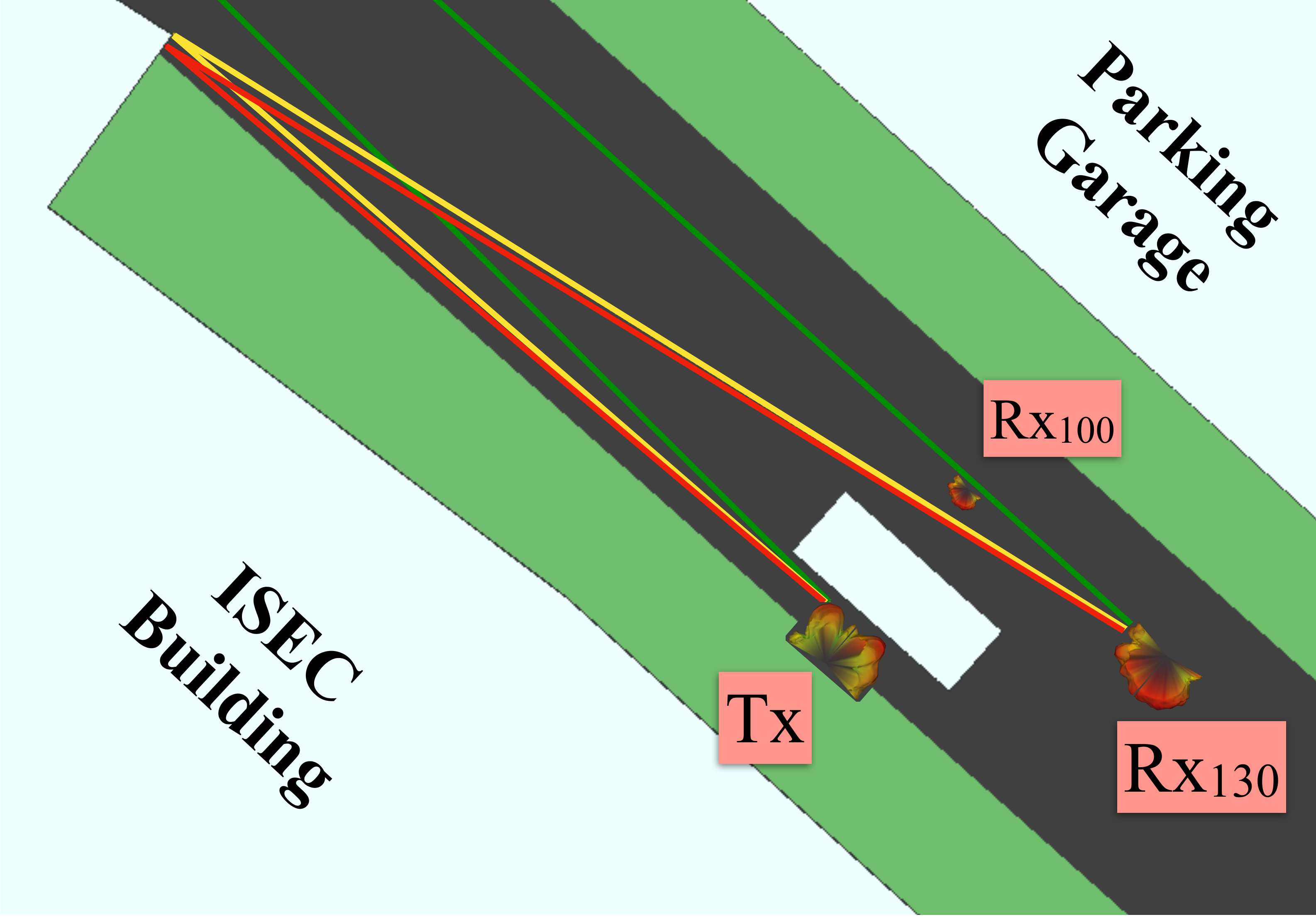}
         \caption{$\twin_{u,2}$ rays for $\env_2$ at Rx$_{130}$.}
         \label{fig:twin2_NLOS_rays}
     \end{subfigure}
          \begin{subfigure}[b]{0.24\textwidth}
         \centering
         \includegraphics[width=\textwidth]{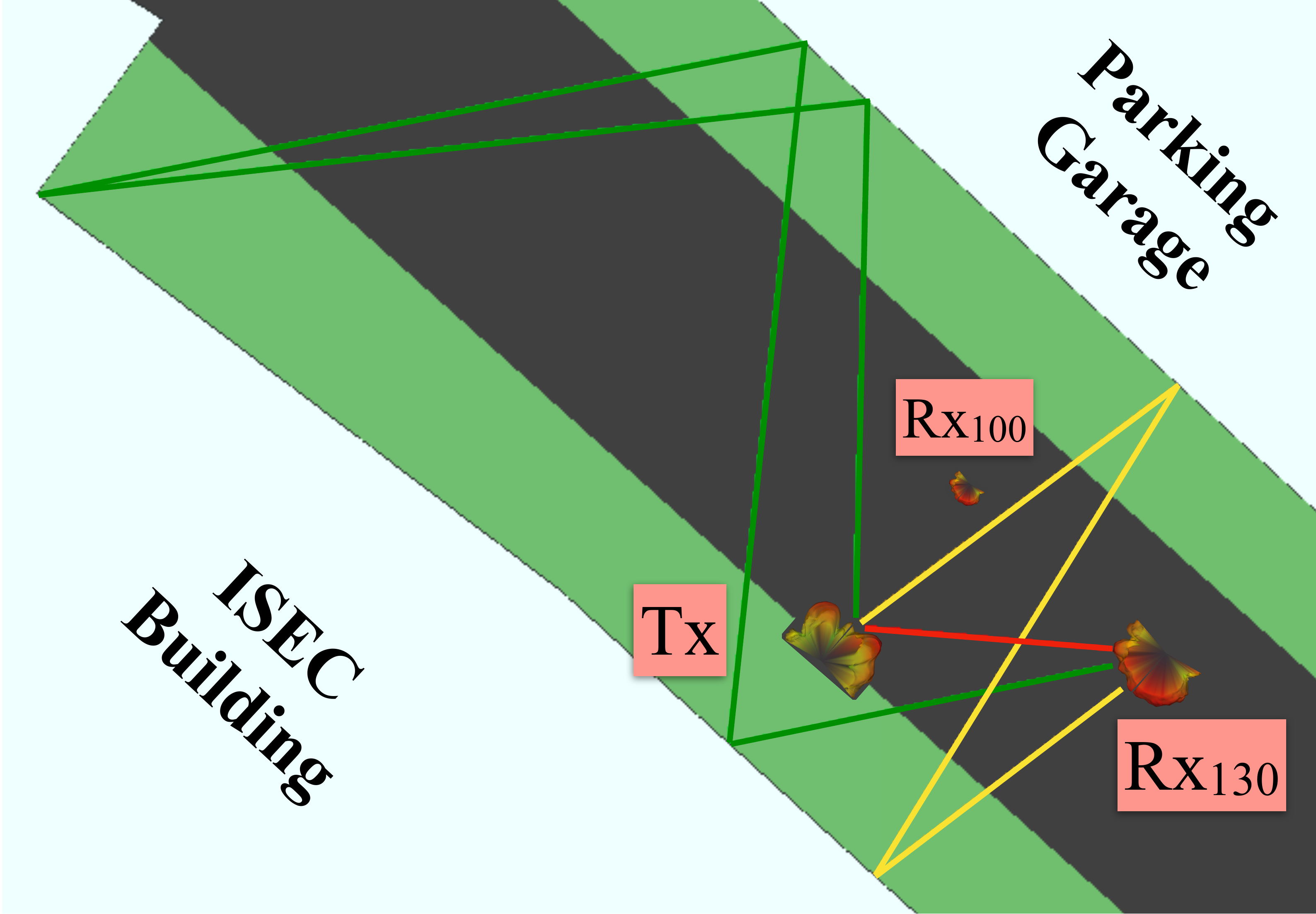}
         \caption{$\twin_{u,3}$ rays for $\env_1$ at Rx$_{130}$.}
         \label{fig:twin3_LOS_rays}
     \end{subfigure}
     \begin{subfigure}[b]{0.24\textwidth}
         \centering
         \includegraphics[width=\textwidth]{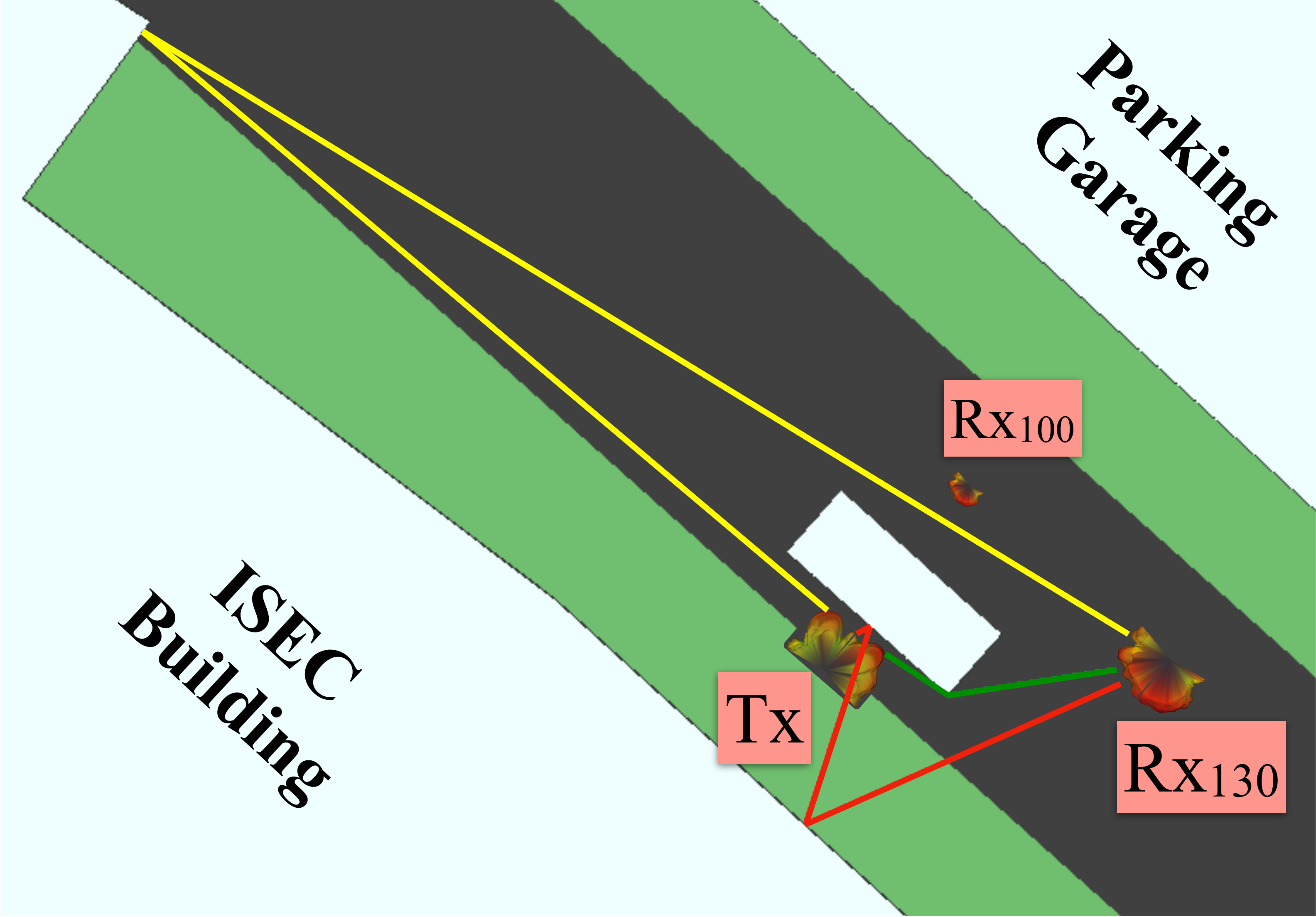}
         \caption{$\twin_{u,3}$ rays for $\env_2$ at Rx$_{130}$.}
         \label{fig:twin3_NLOS_rays}
     \end{subfigure}
     \caption{A set of examples for the first, second, and fifth strongest received rays at Rx\textsubscript{130}, using antenna-10 ($t_{10}$) at the Tx in the $\MV$ $\multiverse$ for $\twin_{u,1}$, $\twin_{u,2}$, and $\twin_{u,3}$ and both $\env_1$ and $\env_2$ scenarios. Red ray is the strongest one, while the ray power decreases towards colder color (red$\rightarrow$yellow$\rightarrow$green). Antenna locations are correct, their patterns are not in scale, enlarged for visual purposes. Due to the mismatch of Tx and Rx antenna height, there was no direct path for LOS rays when the antennas are directly in front of each other.}
     \label{fig:twinsALL_LOS_NLOS_rays}
\end{figure}

\subsection{Observation from Ray Tracing Experiments:}
\label{subsubsec:discuss_wi}
\noindent\textbf{Representative Ray Tracing.}
A collection of outputs from the representative raytracing analysis in the $\WI$ simulator for $\twin_{u,1}$, $\twin_{u,2}$, and $\twin_{u,3}$ are presented in Fig.~\ref{fig:twinsALL_LOS_NLOS_rays} for LOS~($\env_1$) and NLOS~($\env_2$) scenarios. We denote different Rx locations in $\twin_{u,2}$, and $\twin_{u,3}$ with Rx\textsubscript{i}, where $1\leq i \leq |L_{\twin_{u,2}}|, |L_{\twin_{u,3}}|$, respectively, and $|L_{\twin_{u,2}}| = |L_{\twin_{u,3}}| = 200$, according to Tab.~\ref{tab:TwinDiff}. For $\twin_{u,1}$, the only Rx location is denoted as `Rx'~($|L_{\twin_{u,1}}|=1$). In Fig.~\ref{fig:twinsALL_LOS_NLOS_rays}, we present the ray between the Tx and Rx\textsubscript{130} for $\twin_{u,2}$, and $\twin_{u,3}$. Moreover, we show Rx\textsubscript{100} (middle sample point) for location reference. In all examples, Antenna-$10$ ($t_{10}$) is used at Tx. We enlarge the corresponding antenna patterns for visual purposes, but we keep their location precise. In each sub-figure, red ray delivers the highest power and ray strength decreases in the order of red$\rightarrow$yellow$\rightarrow$green, where they represent first, second, and fifth strongest rays.

We observe that when the antennas' $\phi=0^{\circ}$ axes are aligned and the antennas are close, no direct ray is possible. This is due to the height difference between Tx and Rx, which comes from the precise heights we set for the antennas following the FLASH experiment setting~\cite{salehi2022flash}, and the antenna patterns from Talon~\cite{steinmetzer2017compressive}. Thus, the rays in $\twin_{u,1}$, given in Figs.~\ref{fig:twin1_LOS_rays} and~\ref{fig:twin1_NLOS_rays}, are delivered by reflection and diffraction from distant buildings. As a result, the \first~twin, does not feature the reflections from the building in FLASH testbed, making results less reliable. This suggests the idea that a carefully created twin at the same experiment location would yield more robust results, which in fact, we confirm in Sec.~\ref{sec:results}. 
As the antennas become more distant, the height difference is compensated by longer ray travel distance, making a direct ray between Tx and Rx possible~(see red line~\ref{fig:twin2_LOS_rays} and~\ref{fig:twin3_LOS_rays}). In $\twin_{u,2}$ and $\twin_{u,3}$, direct paths are identical, having the $\power_{130}^{10,2} = \power_{130}^{10, 3} = -69.65dBm$. The rest of the LOS rays tend to be drastically different due to the allowed number of reflections ($\rho_2=1$ and $\rho_3=3$). In Fig.~\ref{fig:twin2_LOS_rays} yellow and green rays are delivered to the Rx with at most one reflection, whereas in Fig.~\ref{fig:twin3_LOS_rays} these rays are able to undergo multiple reflections.

In NLOS cases direct path is prevented. Thus, rays have to undergo reflections and/or diffractions in order to reach to the Rx. This is where we see the difference between $\power_{130}^{10,2}$ and $\power_{130}^{10,3}$. The rays in $\twin_{u,2}$ are only allowed to reflect once compared to $\twin_{u,3}$, where $\reflect_i=3$. Thus, in $\twin_{u,2}$, rays are more likely to travel longer to arrive at Rx when there is no direct path. This way, signals in $\twin_{u,2}$ attenuate more, causing lower received power for the strongest beam, e.g., $\power_{130}^{10,2} = -108.4dBm$ and $\power_{130}^{10,3} = -78.16dBm$ in Figs.~\ref{fig:twin2_NLOS_rays} and~\ref{fig:twin3_NLOS_rays}, respectively. Again, due to different number of reflections, we observe yellow and green rays follow different paths in NLOS cases. Overall, we tend to observe $\power_{j}^{m,3} \geq \power_{j}^{m,2}$. Moreover, we expect more accurate beam matching performance for $\twin_{u,3}$, as we also confirm in the Sec.~\ref{sec:results}.

\begin{figure}
     \centering
    \begin{subfigure}[b]{0.24\textwidth}
         \centering
         \includegraphics[width=\textwidth]{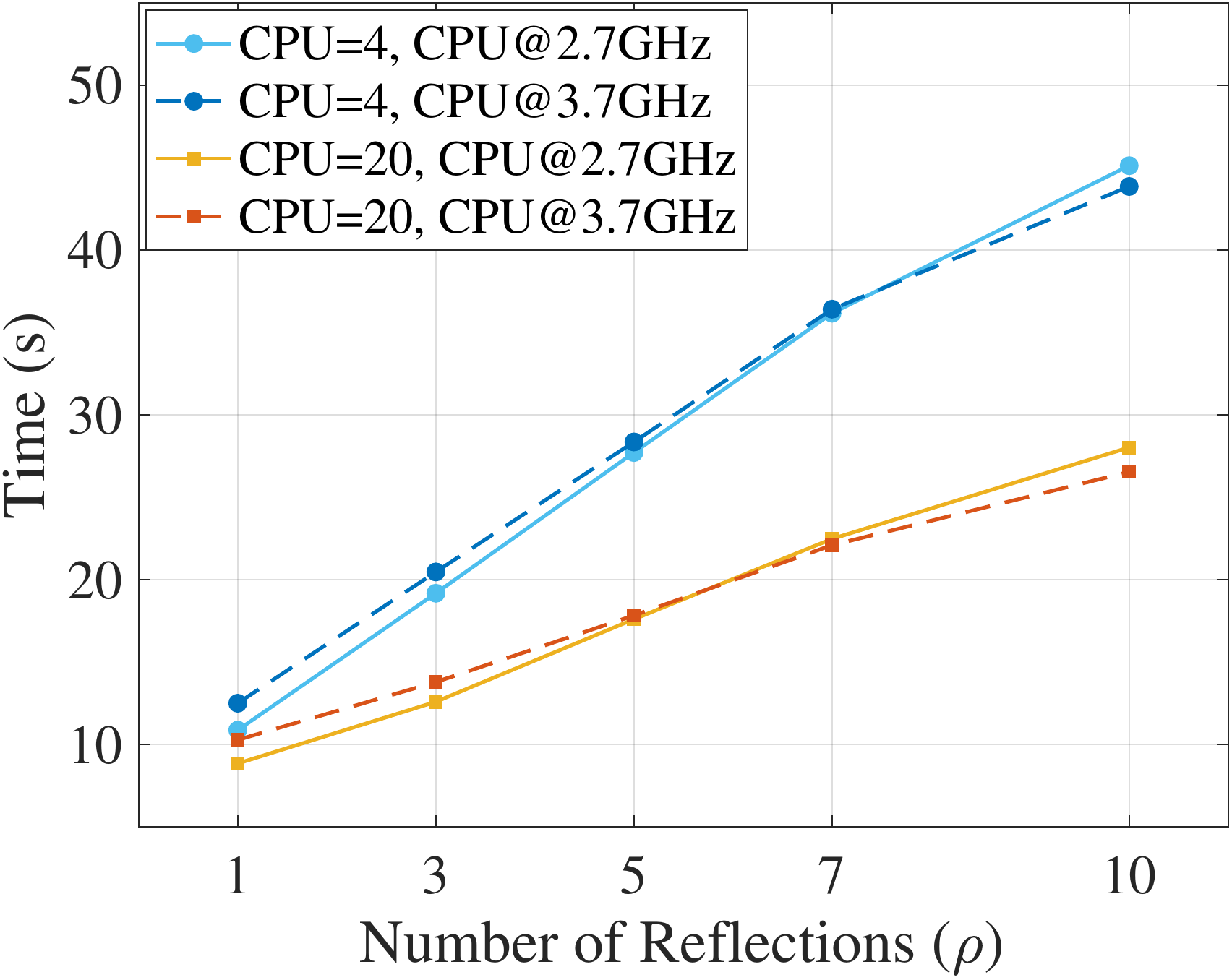}
         \caption{LOS scenarios.}
         \label{fig:CPUtime_LOS}
     \end{subfigure}
     \begin{subfigure}[b]{0.24\textwidth}
         \centering
         \includegraphics[width=\textwidth]{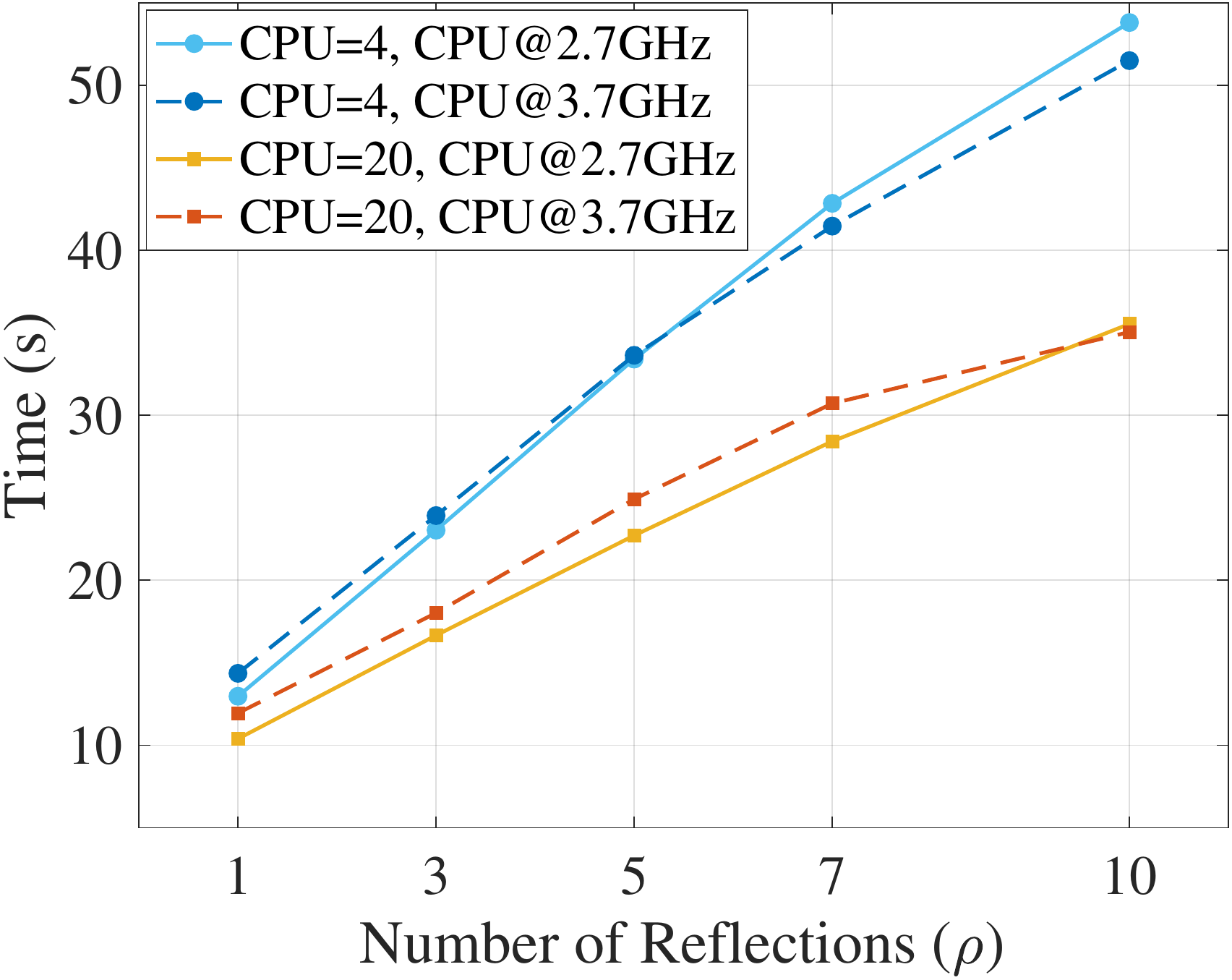}
         \caption{NLOS scenarios.}
         \label{fig:CPUtime_NLOS}
     \end{subfigure}

        \caption{Comparing computation time to run a single antenna experiment on WI for different number of reflections and CPU usage on computers with Intel i9-11900 processor with 32GB RAM (CPU@2.7GHz) and Intel i9-10900K processor with 48GB RAM (CPU@3.7GHz). Overall, NLOS requires more time to complete simulations compared to the LOS case, because the obstacle in the NLOS case creates more surface for rays to reflect, increasing the calculation time. These plots suggest that with more computation power, fast ray tracing and beam selection operations could be performed at a plausible fidelity level.}
     \label{fig:CPUtime}
\end{figure}

\noindent{\bf Cost for Creating the Twins.} The cost of creating $\twin_{u,1}$, $\twin_{u,2}$, and $\twin_{u,3}$ is determined by Eq.~\ref{eqn:cost}. In Eq.~\ref{eqn:cost}, the $\cost^{\twin_{u,i}}_{map}$ consists of one-time cost for: (i) finding appropriate maps and confirming building/object dimensions with their real life counterparts, (ii) antenna pattern design, (iii) precise antenna locations and heights, (iv) obstacle locations and dimensions, (v) deciding on appropriate waveforms and materials, and (vi) selecting the most representative propagation model. We observe $\cost^{\twin_{u,1}}_{map} < \cost^{\twin_{u,2}}_{map} = \cost^{\twin_{u,3}}_{map}$, because $\twin_{u,1}$ is an open area without any building details, whereas $\twin_{u,2}$ and $\twin_{u,3}$ are exact map replica of the real world. On the other hand, $\cost^{\twin_{u,i}}_{lookup}$ is an offline computation cost. In order to create the lookup tables, we use a Dell XPS computer with Intel i9 processor and 32GB RAM following Sec.~\ref{subsubsec:lookup_table}. Detailed analysis of computation time with respect to the CPU usage is given in Fig.~\ref{fig:CPUtime}, where we observe a semi-linear relation between the computation time and the number of reflections~($\reflect$). The $\cost^{\twin_{u,i}}_{lookup}$ for $\twin_{u,1}$, $\twin_{u,2}$, and $\twin_{u,3}$ are given in Tab.~\ref{tab:TwinDiff}.

\begin{table}[t!]
\centering
\begin{tabular}{l|c|c|c|c|c|}
\cline{2-6}
\cline{2-6}
& $\phi$ & $\phi_{\delta}$ & $\theta$ &$\theta_{\delta}$ & SNR \\ \hline
\multicolumn{1}{|l|}{Talon~\cite{steinmetzer2017compressive} } & $[-90^{\circ}, 90^{\circ}]$  &  $1.8^{\circ}$  & $[0^{\circ}, 32.4^{\circ}]$  & $3.6^{\circ}$   & same   \\ \hline
\multicolumn{1}{|l|}{WI}    & $[-100^{\circ}, 100^{\circ}]$  & $2^{\circ}$   & $[0^{\circ}, 36^{\circ}]$  & $4^{\circ}$   & same  \\ \hline
\end{tabular}
  \caption {Comparing Talon and WI metrics.}
  \label{tab:TalonvsWI}
\end{table}

\begin{figure}[t!]
     \centering
    \begin{subfigure}[b]{0.14\textwidth}
         \centering
         \includegraphics[width=\textwidth]
         {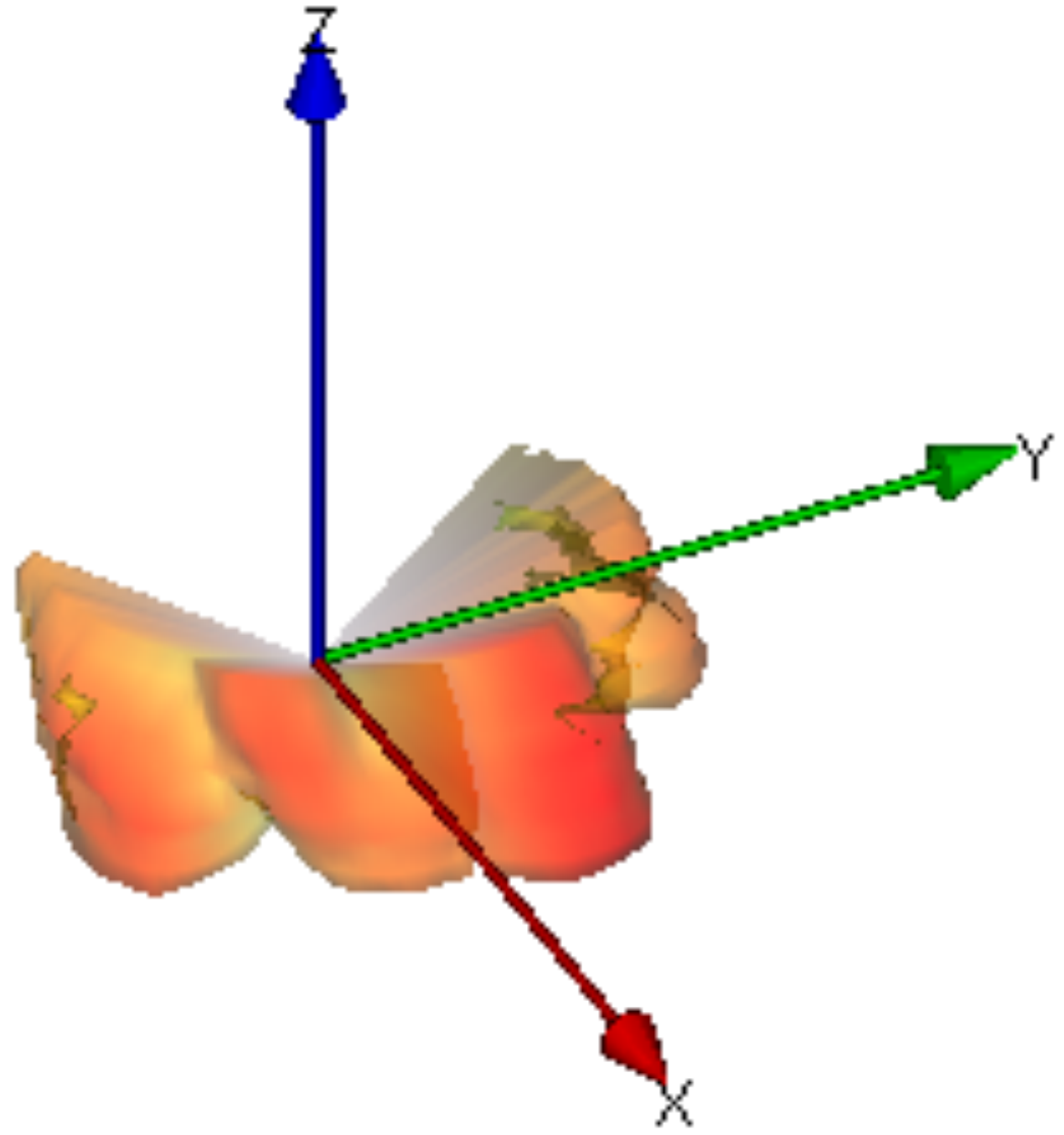}
         \caption{}
         \label{fig:antenna_plot_1}
     \end{subfigure}
     \hspace{5pt}
     \begin{subfigure}[b]{0.14\textwidth}
         \centering
         \includegraphics[width=\textwidth]
         {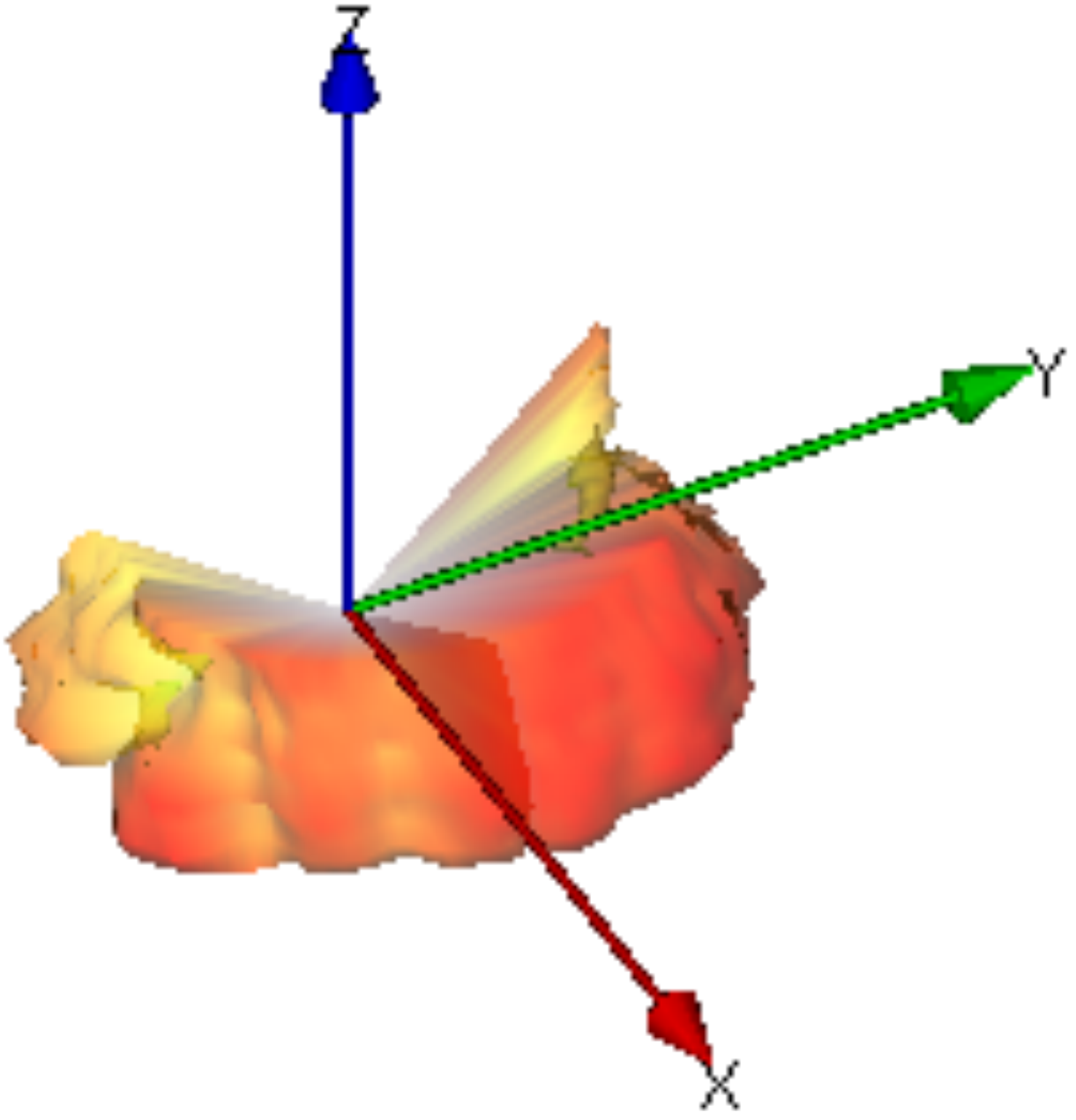}
         \caption{}
         \label{fig:antenna_plot_2}
     \end{subfigure}
     \hspace{24pt}
     \begin{subfigure}[b]{0.15\textwidth}
         \centering
         \includegraphics[width=\textwidth]{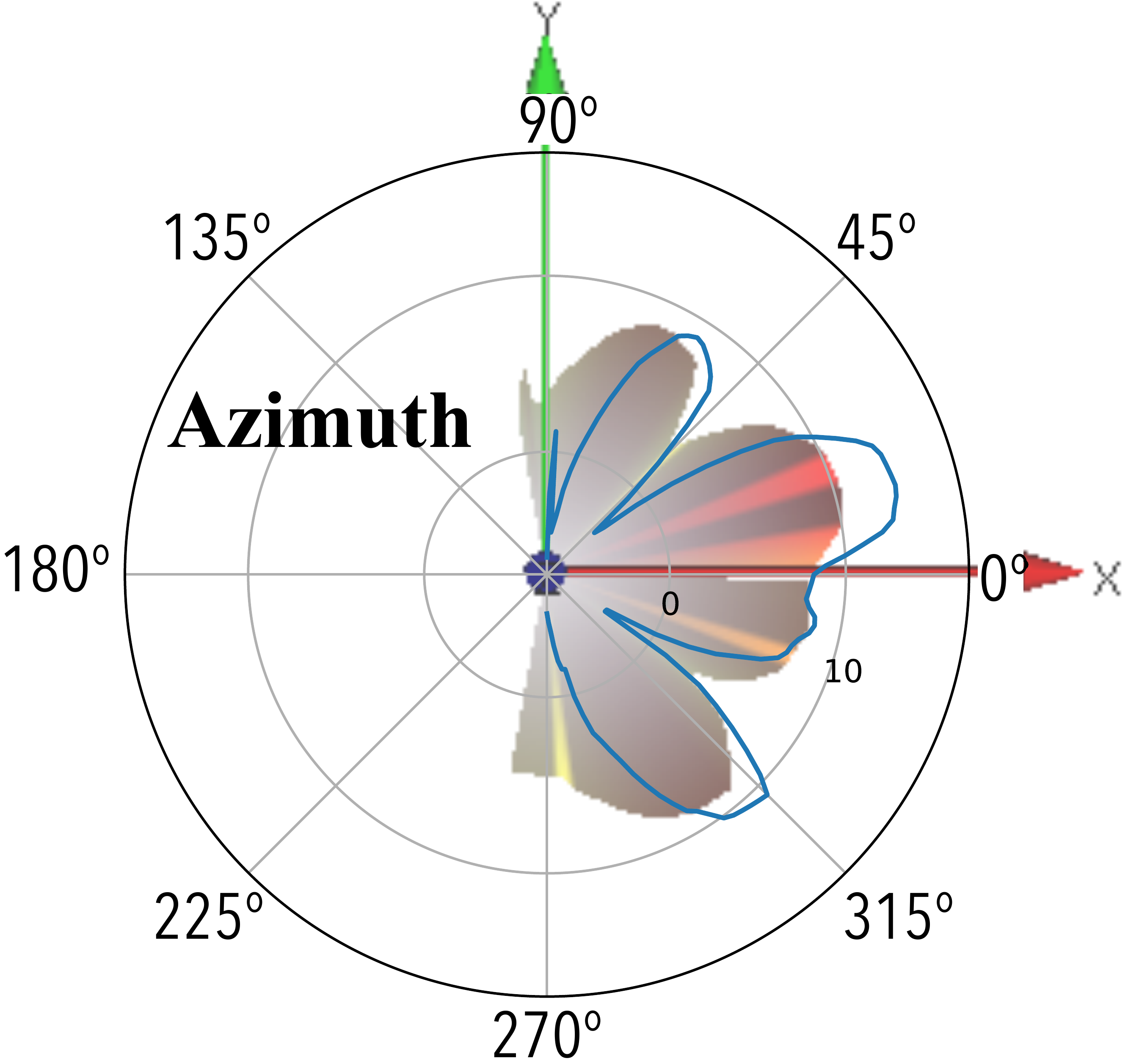}
         \caption{}
         \label{fig:antenna_plot_3}
     \end{subfigure}
     \hfill
     \begin{subfigure}[b]{0.15\textwidth}
         \centering
         \includegraphics[width=\textwidth]{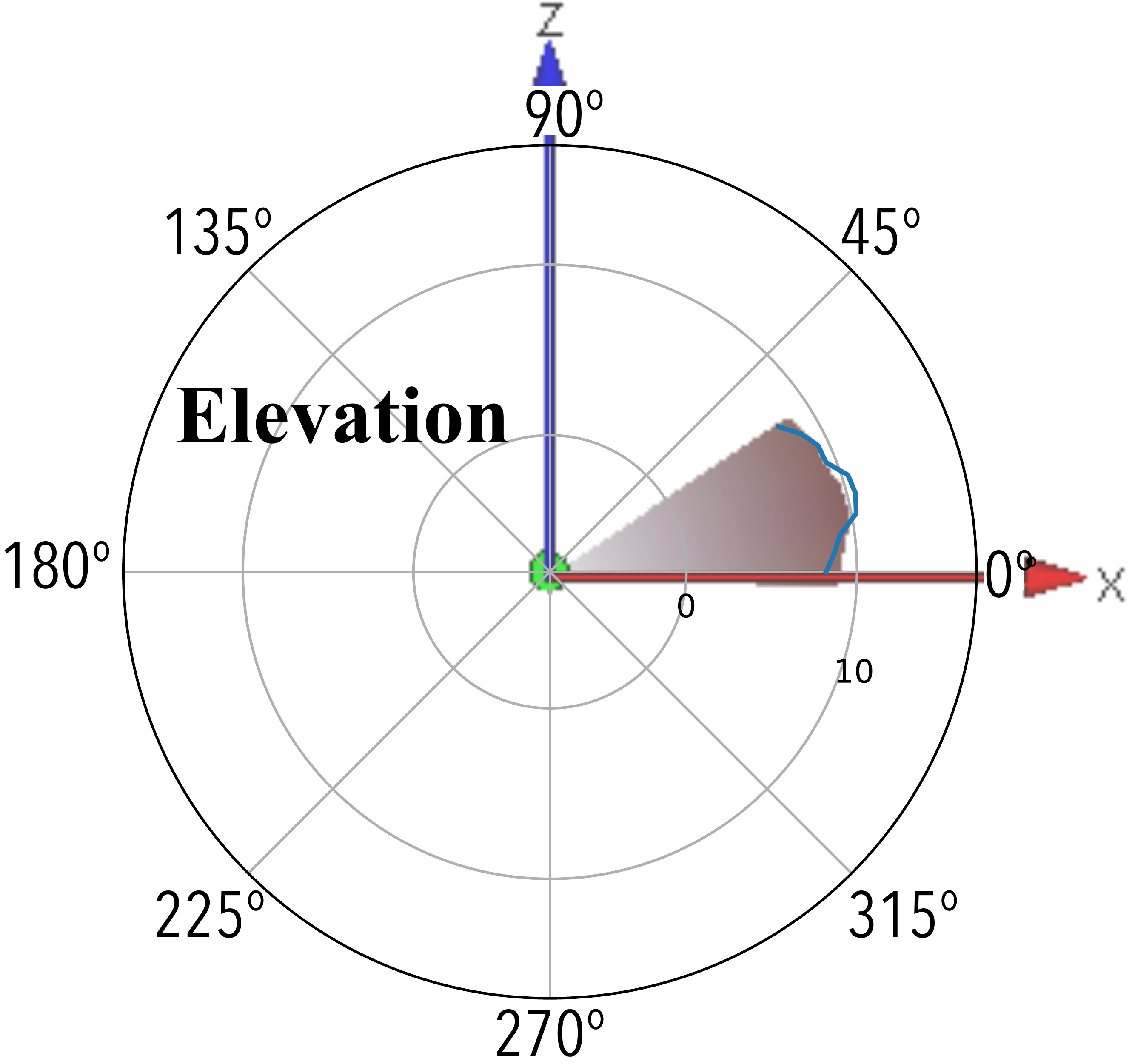}
         \caption{}
         \label{fig:antenna_plot_4}
     \end{subfigure}
     \begin{subfigure}[b]{0.17\textwidth}
         \centering
         \includegraphics[width=\textwidth]{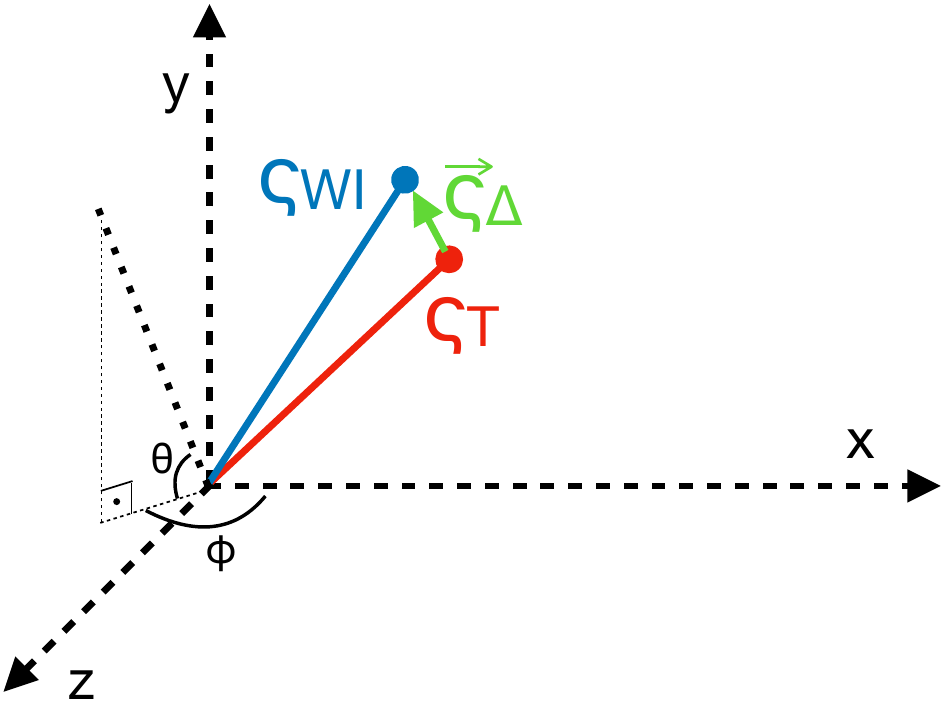}
         \caption{}
         \label{fig:Ant_sim_b}
     \end{subfigure}
     \caption{Re-created antenna pattern examples in Wireless InSite: (a) Tx Antenna-24 ($t_{24}$) and (b) the Rx antenna. Antenna pattern comparisons between the Talon antenna patterns~(solid blue line) and the re-created beams in $\WI$~(shaded area): (c) Azimuth, (d) Elevation. (e) Antenna pattern similarity: The definition of $\phi$ and $\theta$ on the Cartesian coordinates and vector discrepancy calculation, where $\antR_{T}$ and $\antR_{WI}$, are points in experimental and simulation antenna pattern, respectively, and $\vec{\antR}_{\Delta}$ is the vector difference between the $\antR_{T}$ and $\antR_{WI}$.}
     \label{fig:antenna_plots_ALL}
\end{figure}

\noindent {\bf Antenna Similarity (visual representation).}
In Fig.~\ref{fig:antenna_plot_1}-\ref{fig:antenna_plot_4}, we showcase several antenna pattern examples that we have re-created in $\WI$ ray tracing software and provide their comparisons with the Talon antenna patterns for visual similarity. The antenna pattern in Fig.~\ref{fig:antenna_plot_1} belongs to the $24^{th}$ beam~(element $t_{24}$), whereas Fig.~\ref{fig:antenna_plot_2} is the Rx in $\WI$. In Fig.~\ref{fig:antenna_plot_3} and~\ref{fig:antenna_plot_4}, we provide an example comparison between the Talon antenna patterns~(solid blue line) and the re-created beams in $\WI$~(shaded area) for the $24^{th}$ beam~(element $t_{24}$) in 2D azimuth~($\theta = 0^{\circ}$) and elevation~($\phi = 0^{\circ}$), respectively. The slight discrepancy comes from the fact that $\WI$'s user defined antenna patterns only accept sample points with integer increments~($\phi_{\delta}$, $\theta_{\delta}$). To preserve the complete SNR values, we distribute the experimental SNR values over the closest azimuth and elevation regions, while keeping $\phi = 0$ as the reference point; thus, $\phi \in [-100^{\circ}, 100^{\circ}]$ and $\theta \in [0^{\circ}, 36^{\circ}]$. The complete metrics for antenna pattern sample comparison is given in Tab.~\ref{tab:TalonvsWI}.

\noindent {\bf Antenna Similarity (quantitative measures).} In addition to a visual comparison, we also provide a quantitative measure for the antenna similarity between Talon and $\WI$. We measure the antenna discrepancy with an average score from each of these 34 beams, given by the magnitude of the vector differences between experimental and simulation antenna patterns. A visual vector representation of antenna pattern samples is given in Fig.~\ref{fig:Ant_sim_b}, where $\antR_{T}$ and $\antR_{WI}$ represent points in experimental and simulation antenna pattern, respectively, which were created with corresponding SNR values, $\phi$, and $\theta$. The discrepancy score for a single antenna, given in Eq.~\ref{eqn:ant_disc_single}, is calculated by averaging the normalized magnitudes of difference vectors, $\vec{\antR}_{\Delta}$, where the normalization factor is the magnitude of the corresponding antenna pattern sample ($|\vec{\antR}_{WI}|$=$|\vec{\antR}_{T}|$).
\begin{equation}
\Delta_{single} = \frac{1}{Q}\sum_{i=1}^{Q} \frac{|\vec{\antR}_{WI_{i}} - \vec{\antR}_{T_{i}}|}{|\vec{\antR}_{WI_{i}}|},
\label{eqn:ant_disc_single}
\end{equation}
where $Q=1010$~(representing the number of sample points in the antenna patterns), and $0 \leq \Delta_{single} \leq 2$. The final average antenna pattern similarity score, averaged over 34 beams, is found to be $\Delta = 0.0931$ for both scenarios, $\env_1$ and $\env_2$. Smaller the $\Delta$, more similar the antenna patterns are. The designed set of antenna patterns for the $\MV$ using $\WI$ simulation will be published upon acceptance of this paper. 


\section{Performance of the Multiverse}
\label{sec:results}
We validate our proposed framework using the FLASH dataset, presented in Sec.~\ref{sec:dataset_flash}. We consider two set of scenarios, including LOS~($\env_1$) and NLOS~($\env_2$). For each scenario, the $\MV$ consists of three twins each, $\multiverse_1 = \{\twin_{1,1}, \twin_{1,2}, \twin_{1,3}\}$ for LOS scenario as an example. 

\subsection{Evaluation Metrics}
Given the ground-truth measurements from FLASH dataset, $G\in\mathbb{R}^{|\mathcal{B}|}$, and the SNRs $S\in\mathbb{R}^{|\mathcal{B}|}$ for $\mathcal{B}$ beams~(either obtained from the $\MV$ or predicted by DL-based method), we evaluate the beam selection performance as:
\begin{equation}
    \label{eq:top-K-threshold}
    Acc_{(K,T)} = \frac{1}{V} \sum_{l=1}^{V} \id_{(\exists g \subset G |g\in S_k)},\\
\end{equation}
where $V$ denotes the number of samples and $\tau$ is a Boolean predicate, with $\id_{\tau}$ to be 1 if $\tau$ is true, and 0 otherwise. In Eq.~\ref{eq:top-K-threshold}, the set $g$ denotes a subset of beams in G~(ground-truth from FLASH dataset) such that the observed SNR of beams in $g$ is within the $T-$dB threshold of the optimum beam, i.e, $g =\{t_m\in \mathcal{B}| \boldsymbol{SNR}_{t_m}\geq \boldsymbol{SNR}_{t^*}-T\}$, where $\boldsymbol{SNR}_{t^*}$ denotes the SNR of the optimum beam from FLASH ground-truth. Moreover, $S_K$ denotes the top-$K$ beams obtained by either the $\MV$ or DL-based method, defined as:  
\begin{equation}
    \label{eq:top-K-def}
    S_K =\{s| \underset{s\subset \{1,...,|\mathcal{B}|\},|s|=K}{\arg\max}~\sum_{m\in s} SNR_{t_{m}}\},\\
\end{equation}
where $SNR_{t_m}$ denotes the inferred SNR for beam $t_m$. This is intuitive as our observation indicates that there are closely competitive beams in Talon radio~\cite{steinmetzer2017compressive}, used for collecting the FLASH dataset. Thus, selecting any of them results in a favourable observed SNR. For $T=0$ and $K=1$, we get the conventional top-1 accuracy, where only the best beam in FLASH is compared with the best inferred beam. 


\subsection{Motivation for the Multiverse}
\label{sec:motivation_for_multiverse}
We consider a scenario where the mmWave beam initialization is performed by relying on multimodal data and DL-based method proposed in Sec.~\ref{sec:ml_beam_selection}. We follow the models released publicly along with the FLASH dataset to train a CNN~\cite{salehi2022flash}. The proposed architecture is inspired by ResNet~\cite{resnet} and exploits feature level fusion to reinforce the prediction, by incorporating the sensor data from all modalities. In Tab.~\ref{tab:motivation_multivers}, we benchmark the performance of the DL-based method when the CNN is trained on one scenario and tested on another~(unseen during training), e.g. trained on LOS scenario $\env_1$ and tested on NLOS scenario $\env_2$. 
We observe that the maximum drop in accuracy~($Acc_{(1,0)}$) is $73.42\%$ while making a transition from $\env_2$ to $\env_1$  scenarios. Thus, we conclude that an DL-only method fails when unseen environments are encountered.
\begin{table}[t]
\centering
\resizebox{\linewidth}{!}{
    \begin{tabular}{||c|c|c|c|c||} 
    \hline
    \multirow{2}{1.70cm}{Train Scenario}&\multirow{2}{1.8cm}{Train $Acc_{(1,0)}$} &\multicolumn{2}{c|}{Test $Acc_{(1,0)}$} &\multirow{2}{2.1cm}{Drop in $Acc_{(1,0)}$}\\ 
    \cline{3-4}
    & & $\env_1$ & $\env_2$ &  \\\hline\hline
    $\env_1$ (LOS) & $83.9$ & $66.40$ & $10.30 $&$\mathbf{56.1}$\\\hline
    $\env_2$ (NLOS) & $90.97$ & $5.96$ & $79.38$ &$\mathbf{73.42}$\\\hline
    \end{tabular}}
    \caption{The $Acc_{(1,0)}$ while being exposed to the unseen environments. The DL-based methods experience extensive drop in accuracy due non-adaptability to the changes in the environment.}
    \label{tab:motivation_multivers}
\end{table}
\subsection{Validation of the Multiverse Concept}
\label{sec:ValMultiverse}

\begin{figure}[t!]
\begin{subfigure}{\linewidth}
  \centering
\includegraphics[width=0.7\linewidth]{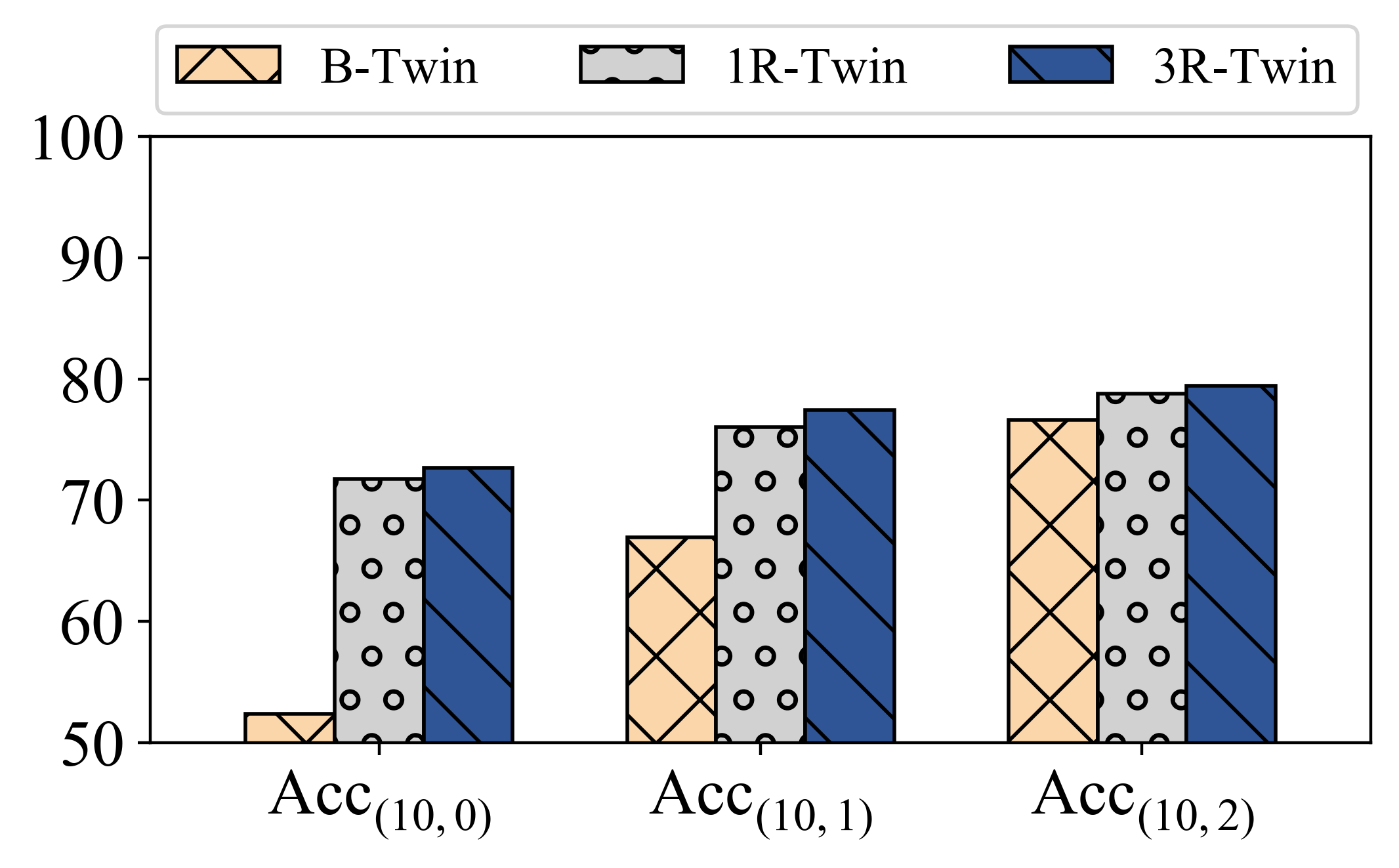}  
  \caption{}
  \label{fig:validation_WI_LOS}
\end{subfigure}
\begin{subfigure}{\linewidth}
  \centering
  \includegraphics[width=0.7\linewidth]{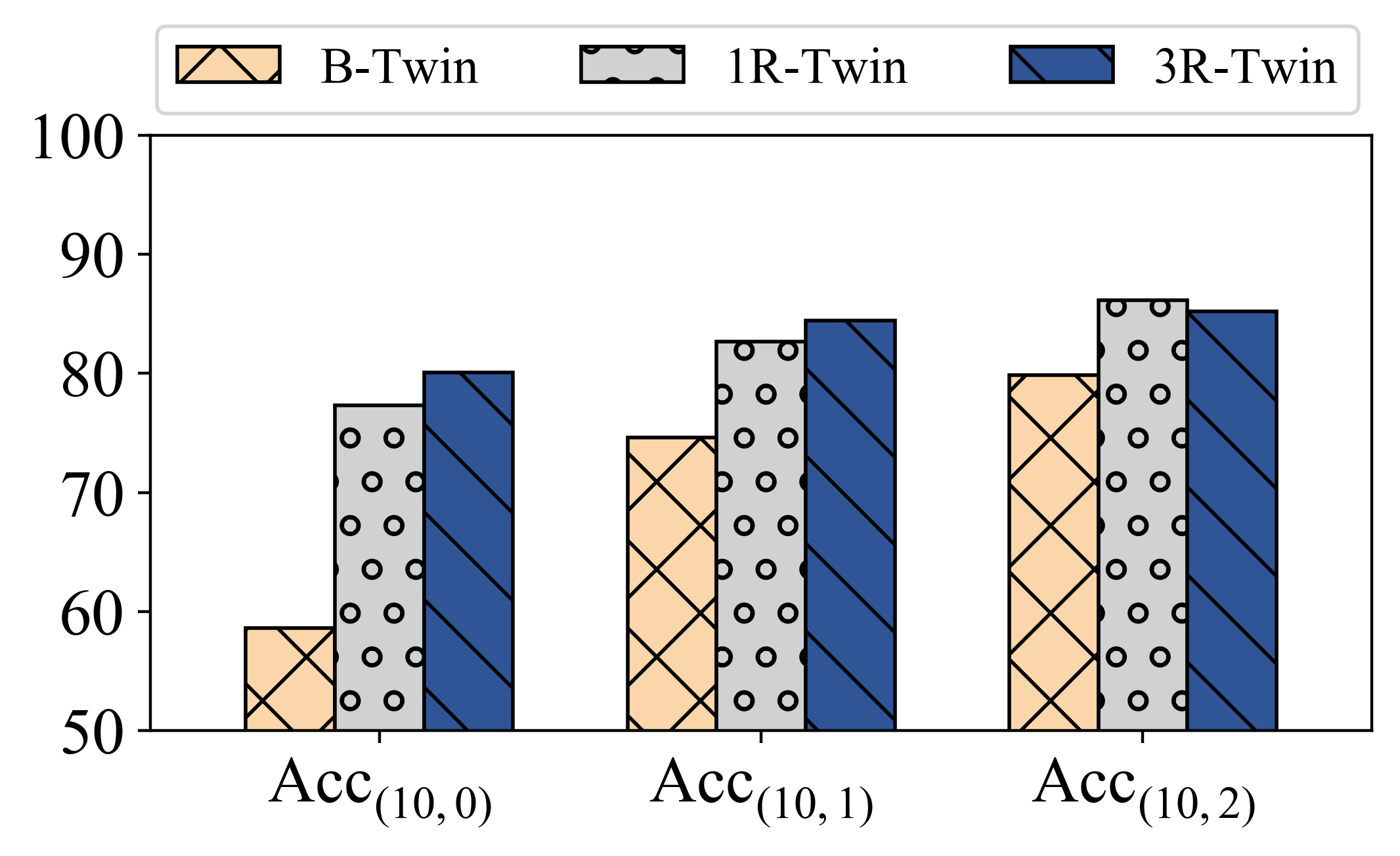}  
  \caption{}
  \label{fig:validation_WI_NLOS}
\end{subfigure}
\caption{Benchmarking the performance of three twins in the $\MV$ against the real-world measurements for (a) LOS $\env_1$ and (b) NLOS $\env_2$ scenarios. The accuracy for $K=10$ and SNR thresholds of $0,1,2$ dB indicate the fidelity of the $\MV$ to the real world setting. The $\third$ twin and $\second$ twin exhibits more fidelity over the $\first$ twin.}
\label{fig:DT_LOS_NLOS}
\end{figure}

In Fig.~\ref{fig:DT_LOS_NLOS}, we compare the performance of the $\MV$ against the ground-truth measurements from the FLASH dataset to validate the fidelity of emulation outputs. We report the metric $Acc_{(K,T)}$ in Eq.~\ref{eq:top-K-threshold} and set the parameter $K$ as 10 and gradually relax the SNR threshold $T$ with $T=\{0\text{dB},1\text{dB},2\text{dB}\}$. Note that $T=0$ is most extreme case where only the optimum beam in FLASH is considered for justification. From Fig.~\ref{fig:DT_LOS_NLOS}, we observe that for the LOS scenario~($\env_1$), \first~twin $\twin_{1,1}$ provides the $Acc_{(10,0)}$ of $52.38\%$, while \second~and~\third~twins $\twin_{1,2}$ and $\twin_{1,3}$ perform closely with accuracy of $71.74\%$ and $72.69\%$, respectively. The $Acc_{(10,0)}$ increases as we relax the threshold on SNR and ranges between 76-79\% across all three twins. For NLOS scenario $\env_2$, we observe that \third~twin $\twin_{2,3}$ exhibits superiority in $Acc_{(10,0)}$ against $\twin_{2,1}$ and $\twin_{2,2}$ by $21.49\%$ and $2.78\%$, respectively. Overall, the designed $\MV$ depicts the accuracy ranging between $52.38-80.08\%$, $66.92-84.41\%$ and $76.63-85.22\%$ for SNR thresholds of 0, 1 and 2dB, respectively, for both LOS~($\env_1$) and NLOS~($\env_2$) scenarios across all twins.

\subsection{Comparing the Performance of Twins}
\label{sec:compareTwins}
\begin{figure*}[ht]
\begin{subfigure}{\linewidth}
  \centering
\includegraphics[width=0.9\linewidth]{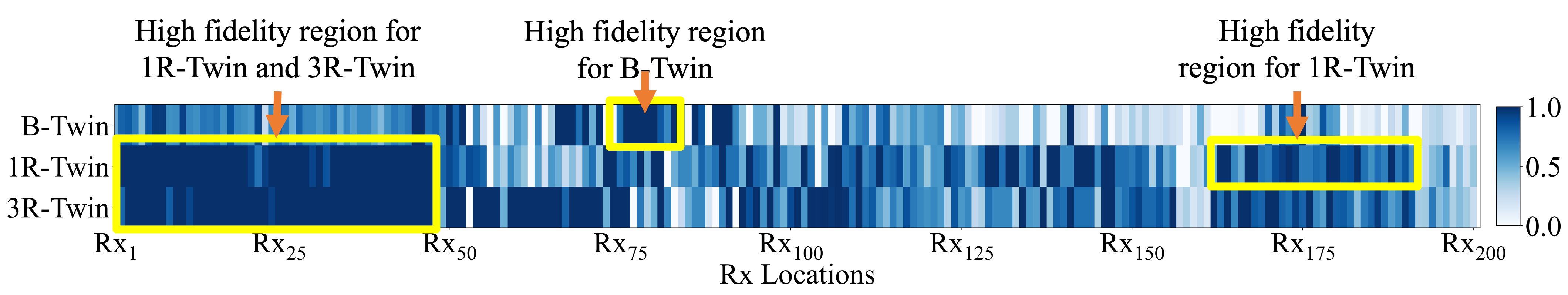}  
  \caption{}
  \label{fig:heatmap_twins_los}
\end{subfigure}
\begin{subfigure}{\linewidth}
  \centering
  \includegraphics[width=0.9\linewidth]{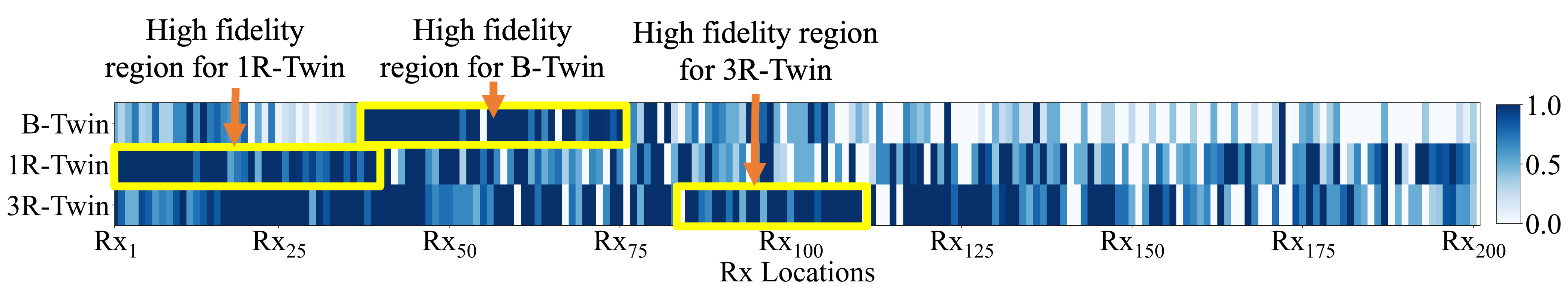}  
  \caption{}
  \label{fig:heatmap_twins_nlos}
\end{subfigure}
\caption{Probability of inclusion for three twins for (a) LOS $\env_1$ and (b) NLOS $\env_2$ scenarios. Each column shows a region on the road and each row depicts the probability of inclusion for one of the twins. While the performance on twins varies in different regions, the $\third$ twin outperforms the $\first$ and $\second$ twins on average. All twins offer higher probability of inclusion in $\env_1$ compared to $\env_2$. A few sample regions where one twin has more fidelity than other, is highlighted in both (a) and (b).}
\label{fig:comapre_twins}
\end{figure*}

In Fig.~\ref{fig:comapre_twins}, we compare the probability of inclusion $p(K,LT (\twin_{\unlabeled,i}),\env_\unlabeled,r)$ for $K=10$, three twins~($i=\{1,2,3\}$), and both LOS and NLOS scenarios~($\unlabeled=\{1,2\}$). In this figure, each cube represents the location on the road $r$ for which the probability of inclusion is calculated, with $20cm$ spacing, as explained in Sec.~\ref{sec:multiverse_creation}. From this figure we conclude three observations. First, the performance of the twins depends on the location of the receiver on the road. This is intuitive as the number of reflections in real world might vary in each location, unlike the $\MV$ where it is constant for each twin. As a result, each twin might be closer to the real world scenario on a case by case basis. This is also in support to the idea of region wise selection of twins in Eq.~\ref{eq:opt_twin_selection}. Second, we observe that $\third$ twin performs better than $\first$ and $\second$ with the average probability of inclusion of $74.87-78.68\%$ compared to that of $43.90-47.48\%$ and $66.43-76.60\%$ for LOS~($\env_1$) and NLOS~($\env_2$) scenarios, respectively. Third, the $\MV$ provides higher probability of inclusion in $\env_1$ compared to $\env_2$ by $4-8\%$ margin. 
\subsection{Digital Twin Selection from the Multiverse}
\begin{figure*}[ht]
\begin{subfigure}{0.35\textwidth}
  \centering
  \includegraphics[width=\linewidth]{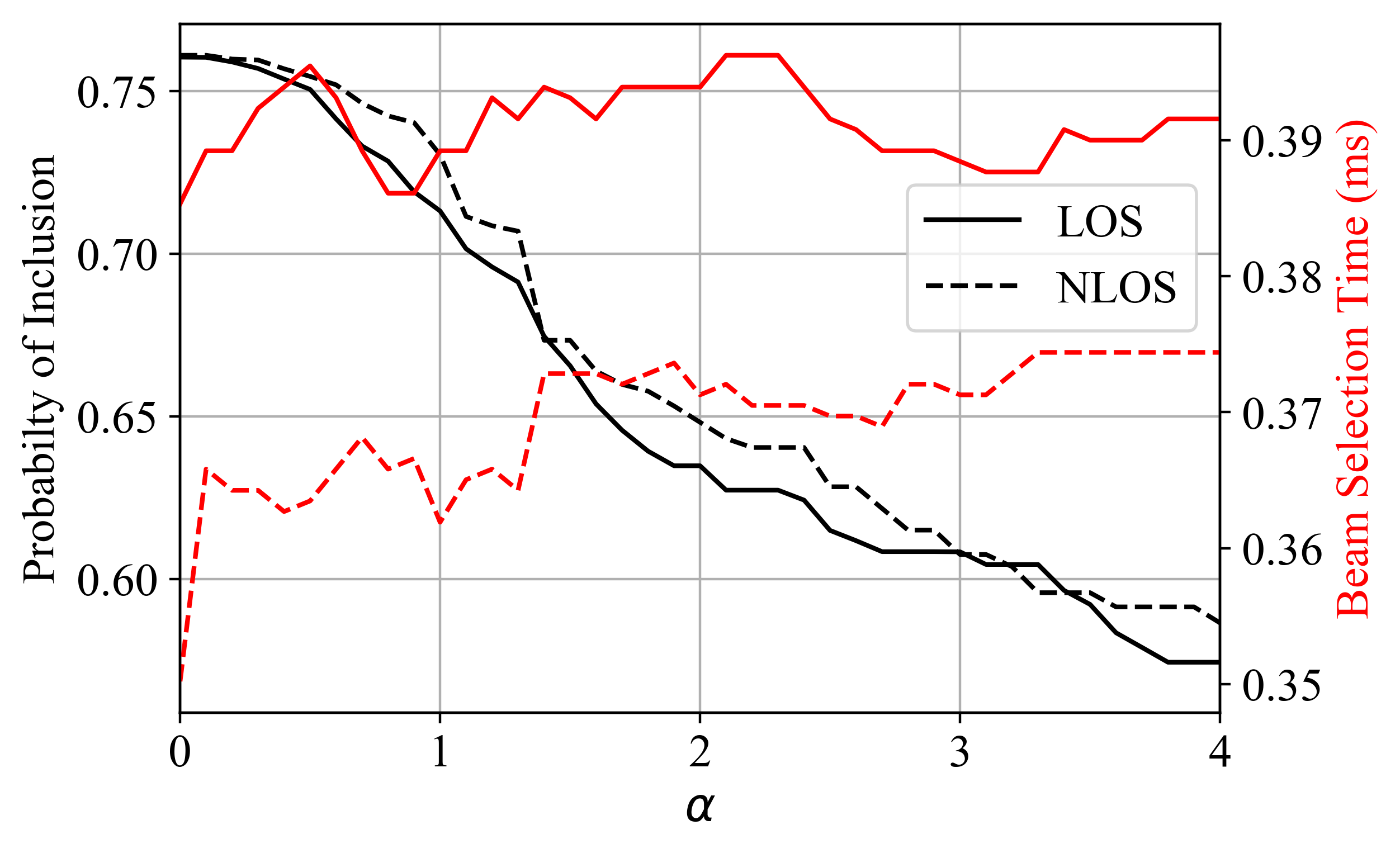}  
  \caption{}
  \label{fig:plot_alpha}
\end{subfigure}
\hspace{2mm}
\begin{subfigure}{0.31\textwidth}
  \centering
  \includegraphics[width=\linewidth]{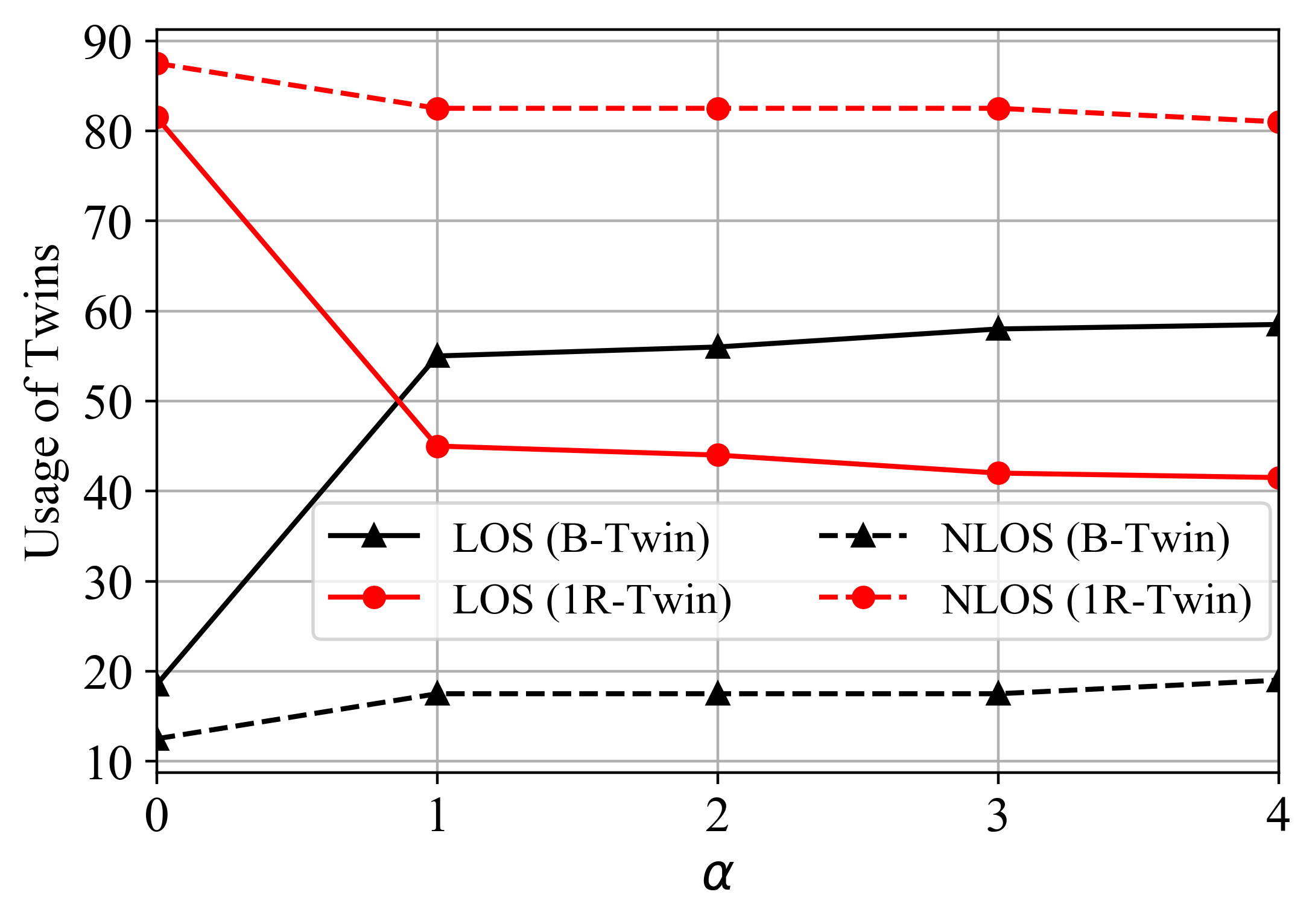}  
  \caption{}
  \label{fig:computation_constraint}
\end{subfigure}
\hspace{2mm}
\begin{subfigure}{0.31\textwidth}
  \centering
  \includegraphics[width=\linewidth]{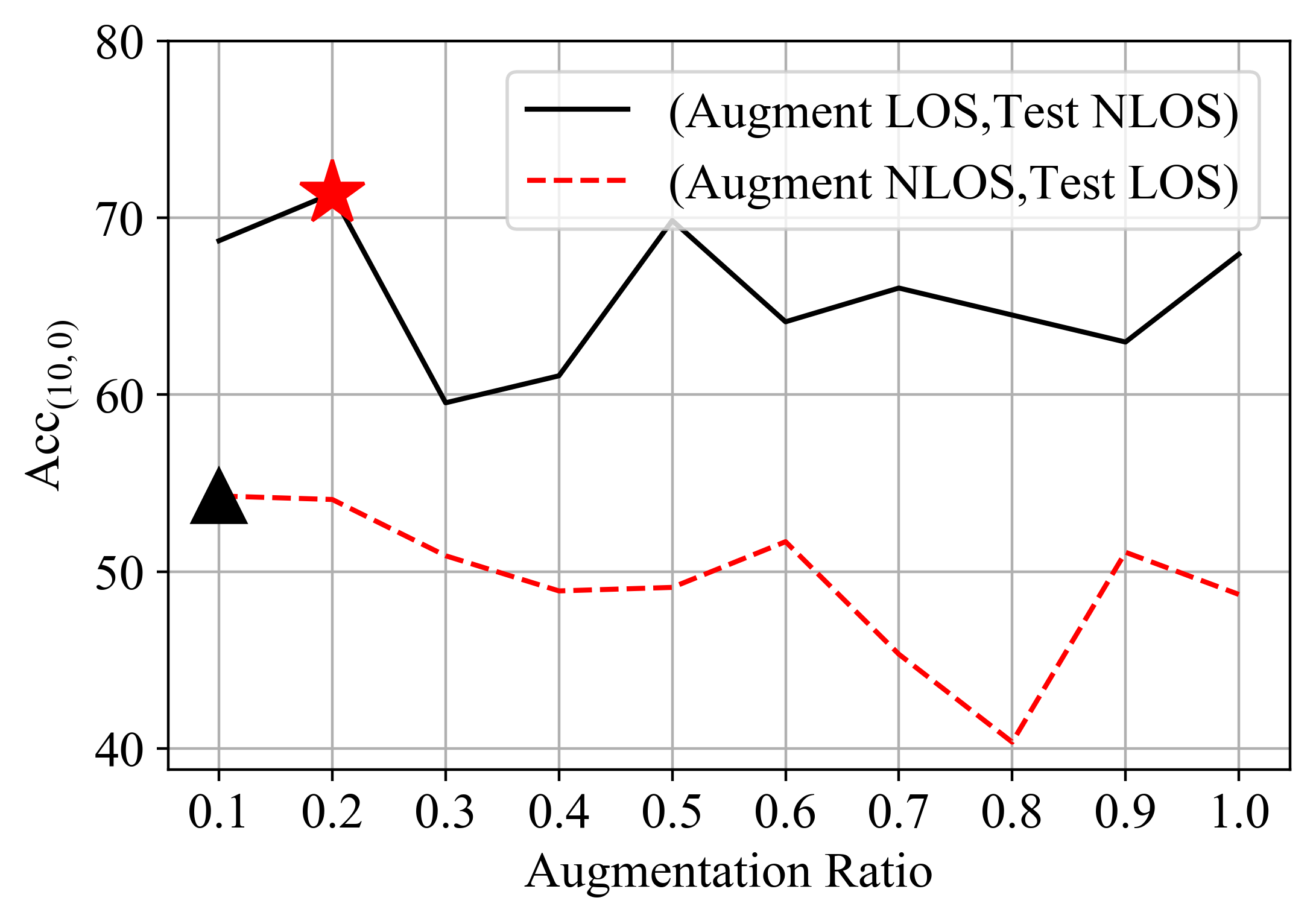}  
  \caption{}
  \label{fig:augmentaion_exp}
\end{subfigure}
\caption{(a) Analysis of inclusion probability and beam selection time for different $\alpha$ values in Eq.~\ref{eq:opt_twin_selection} when the $\cost_{Comm}$ is set to be less than $1ms$. The probability of inclusion decreases as the optimization problem is weighted on the second term in Eq.~\ref{eq:opt_twin_selection} that encourages minimizing the computation and communication cost. However, the beam selection time fluctuates as a result of alternating between choosing lower $K$ or a twin with higher fidelity. (b) The usage of twins when the computation constraints rules out \third~twin in Eq.~\ref{eq:opt_twin_selection}, which has the highest computation cost. When $\alpha=0$, $\second$ twin is used by 81.5\% while $\first$ twin is selected by 18.5\%. When $\alpha$ increases $\first$ twin is used more often to decrease the computation and communication time. (c) The $Acc_{(10,0)}$ of DL-based method when fine-tuned by the labels obtained by $\third$ twin. The $Acc_{(10,0)}$ with fine-tuning is bounded by $Acc_{(1,0)}$ and $Acc_{(10,0)}$ in the $\MV$. The {\em max} values are marked in the figure.}
\label{fig:result_three}
\end{figure*}

\noindent{\bf Effect of Control Parameter.} The proposed optimization problem in Eq.~\ref{eq:opt_twin_selection} selects the optimum twin according to the constraints on communication and computation latency. In Fig.~\ref{fig:plot_alpha}, we show the effect of control parameter $\alpha$ on the optimization problem in Eq.~\ref{eq:opt_twin_selection} for LOS~($\env_1$) and NLOS~($\env_2$) scenarios. We observe that by increasing $\alpha$ the probability of inclusion decreases, while we observe a fluctuating pattern in beam selection time. Note, that by increasing $\alpha$ the optimization problem in Eq.~\ref{eq:opt_twin_selection} is enforced to decrease $K$ and select the twins with lower communication and computation cost. On the other hand, decreasing $K$ motivates choosing twins with higher probability of inclusion to maximize the objective. The monotonic decrease in probability of inclusion in Fig.~\ref{fig:plot_alpha} indicates that this metric is mostly effected by $K$ and is less sensitive to the twins. On the other hand, the beam selection time is affected by both $K$ and twins which results in fluctuations observed in Fig.~\ref{fig:plot_alpha}, as $\alpha$ changes.

\noindent{\bf Effect of Latency Constraint.}
In Tab.~\ref{tab:usage_per_twin}, we report the percentage of the times that each of the twins were selected under three communication latency constraints, $\cost_{{Comm}_1}=0.78ms$, $\cost_{{Comm}_2}=1.56ms$ and $\cost_{{Comm}_3}=2.34ms$. Moreover, we report the selection percentage for $\alpha=\{0,0.2\}$. For $\alpha=0$, we observe, as the constraint on communication latency is relaxed, the percentage of the times that the twin with higher complexity~(and higher probability of inclusion) is selected increases. In particular, the usage for \second~twin decreases from $51\%$ and $33\%$ to $26\%$ and $25.5\%$ for LOS and NLOS scenarios, respectively. On the other hand, the usage for \third~twin increases by $23\%$ and $4\%$ for each scenario. This is intuitive as relaxing the communication latency constraint indicates that the users are inclined to have a more robust than fast communication; thus, the twin with higher complexity and higher probability of inclusion is more likely to be chosen. For \first~twin, we do not observe significant change in usage of each twin and a range of $4.5-8.5\%$ for all communication latency constraints and both LOS~($\env_1$) and NLOS~($\env_2$) scenarios.

For $\alpha=0.2$, we observe that although the communication latency is relaxed, the usage for \second~and \third~twins  decreases, while the usage for the \first~twin increases. In particular, we observe that in this case, usage of \first~twin increases by $40-45.5\%$ while the usage of \second~and \third twins decreases by $26.5-37.5\%$ and $5-13.5\%$, respectively. Note that by relaxing the communication latency constraint, the probability of inclusion increases. On the other hand, by imposing the second term in Eq.~\ref{eq:opt_twin_selection} with higher weights, the effect of the second term is more pronounced. Thus, the optimization problem is more likely to choose \first~twin with less computation time, despite relaxing the communication constraint.


\begin{table*}[hbtp]
    \centering
\resizebox{0.65\linewidth}{!}{
    \begin{tabular}{||c|c|c|c|c|c|c||} 
    \hline
    \multirow{2}{1cm}{Twin}&\multicolumn{3}{|c|}{LOS~$\env_1$~($\alpha$=0)}& \multicolumn{3}{|c|}{LOS~$\env_1$~($\alpha$=0.2)}\\\cline{2-7} 
    & $\cost_{{Comm}_1}$ &$\cost_{{Comm}_2}$ & $\cost_{{Comm}_3}$& $\cost_{{Comm}_1}$ & $\cost_{{Comm}_2}$ & $\cost_{{Comm}_3}$\\\hline\hline
    B-Twin ($\twin_{1,1}$) & $6.5$ & $4.5$ & $8.5$ &  $35.5$ & $61.5$ &$78$ \\
    \hline
    1R-Twin ($\twin_{1,2}$) & $51$ & $32.5$ & $26$ & $57.5$ & $36.0$ &$20$ \\ \hline
    3R-Twin ($\twin_{1,3}$) &  $42.5$ & $63$ & $65.5$ & $7$ & $2.5$  & $2$ \\ \hline\hline\hline

    \multirow{2}{1cm}{Twin}&\multicolumn{3}{|c|}{NLOS~$\env_2$~($\alpha$=0)}& \multicolumn{3}{|c|}{NLOS~$\env_2$~($\alpha$=0.2)}\\\cline{2-7} 
    & $\cost_{{Comm}_1}$ &$\cost_{{Comm}_2}$ & $\cost_{{Comm}_3}$& $\cost_{{Comm}_1}$ & $\cost_{{Comm}_2}$ & $\cost_{{Comm}_3}$\\\hline\hline
    B-Twin ($\twin_{2,1}$) & $5$ & $7$ & $8.5$ & $27$ &$52$  &$67$\\
    \hline
    1R-Twin ($\twin_{2,2}$) &$33$  & $27.5$& $25.5$ & $54.5$ & $38.5$& $28$\\ \hline
    3R-Twin ($\twin_{2,3}$) & $62$ & $65.5$&  $66$ & $18.5$ &$9.5$ &$5$\\ \hline 
    
    \end{tabular}}
    \caption{The breakdown of the usage (in $\%$) for three twins in the $\MV$ for LOS~($\env_1$) and NLOS~($\env_2$) scenarios while imposed with three communication latency constraints and $\alpha$=\{$0$, $0.2$\}. When $\alpha=0$, relaxing the communication constraint results in an increase in choosing $\third$ twin and decrease in choosing $\first$ and $\second$ twins. However, when $\alpha \neq 0$, the urge to select the twin with lower communication and computation complexity leads to increasing the usage of $\first$ twin with minimum computation complexity. }
    \label{tab:usage_per_twin}
\end{table*}

\noindent{\bf Effect of Computation Constraint.} We note that the computation complexity of three twins in the $\MV$ is related as $\twin_{u,1}<\twin_{u,2}<\twin_{u,3}$, regardless of the scenarios~($\forall u$). In this experiment, we consider a case in which \third~twin is not a viable option due to computation constraints on the edge. Moreover, we fix the communication constraint to be $\cost_{{Comm}_1}=0.78ms$. In Fig.~\ref{fig:computation_constraint}, we report the model usage versus $\alpha$ for $\twin_{u,1}$ and $\twin_{u,2}$ and both scenarios $u=\{1,2\}$. We observe that for $\alpha=0$ the usage for \first~and \second~twins is $18.5\%$ and $81.5\%$ for LOS scenario~($\env_1$) and $12.5\%$, $87.5\%$ for NLOS scenario~($\env_2$), respectively. Moreover, we observe as $\alpha$ increases, the optimization problem in Eq.~\ref{eq:opt_twin_selection} is choosing \first~twin more often. 

\subsection{Fine-Tuning with the Multiverse Ground-Truth}
We consider a scenario in which a percentage of labels from the $\MV$~(ray tracing outputs), which we refer to as \textit{labeling ratio}, is paired with the local multimodal sensor data to fine-tune the local model at the vehicle. For testing, we only use the labels from FLASH dataset that is our ground-truth. We generate the labels from \third~twin in the $\MV$ which offers the highest probability of inclusion. In one experiment, we use the multimodal sensor data and labels from FLASH ground-truth for $\env_1$. We then label the sensor data within $\env_2$ using the ray tracing outputs from the $\MV$~(not FLASH) and fine-tune the model. We then repeat the same experiment with $\env_1$ and $\env_2$ role-reversed. As Sec.~\ref{sec:motivation_for_multiverse}, we use the model architecture released by FLASH framework~\cite{salehi2022flash}. In Fig.~\ref{fig:augmentaion_exp}, we report the $Acc_{(10,0)}$ for both cases. We note that the $Acc_{(1,0)}$ and $Acc_{(10,0)}$ for LOS~($\env_1$) and NLOS~($\env_2$) scenarios in the $\MV$ range between $41-72\%$ and $47-80\%$, respectively. On the other hand, we observe that $Acc_{(10,0)}$ with fine-tuning ranges between $40.35-54.27\%$ and $59.54-69.84\%$ for $\env_1$ and $\env_2$. Interestingly, the $Acc_{(10,0)}$ with fine-tuning is within the $Acc_{(1,0)}$ and $Acc_{(10,0)}$ from the $\MV$. We note that the labeling strategy in FLASH framework in one-hot encoding; thus, it is expected for the accuracy with fine-tuning to be higher than $Acc_{(1,0)}$ from the $\MV$. On the other hand, since the labels are synthetically generated, the fine-tuning  accuracy is bounded by $Acc_{(10,0)}$ from the $\MV$. We observe the best performance for labeling ratio of $0.1$ and $0.2$ for $\env_1$ and $\env_2$, respectively, marked with star and triangle in Fig.~\ref{fig:augmentaion_exp}.
\subsection{Multiverse against Exhaustive Search and State-of-the-art}
\label{sec:results_compare}
\noindent
$\bullet$ {\bf Comparing with Exhaustive Search: }
In the $\MV$, a look up table for each twin is sent in the downlink. This look up table includes information about expected SNR of each beam across all Rx locations and the computation cost for each twin. Whenever, the DL-based method identified as faulty, the vehicle runs the optimization problem in Eq.~\ref{eq:opt_twin_selection} to obtain the optimum twin and associated top-$K$ beams according to the computation and communication constraints. We note that the aforementioned look up tables for each twin occupy maximum $\sim28KB$ per twin. Moreover, the Talon AD7200 router offer communication in both mmWave and 5~GHz bands. We consider the 5~GHz band as the control channel that provides data rate of $1733Mbps$~\cite{TalonSpec_website}. As a result, sending the look up tables in downlink takes $\sim0.1292ms$. On the other hand, the local execution of Eq.~\ref{eq:opt_twin_selection} takes $\sim 23\mu s$ on a commercial laptop. Moreover, we note that the selected top-$K$ beams in the $\MV$ are bounded by 12 while targeting the inclusion probability of $97\%$ (there is no need to sweep more than 12 beams with the $\MV$). This corresponds to beam sweeping times of $0.0373ms$~($K=1$) and $0.4482 ms$~($K=12$), respectively. Thus, we conclude that the end-to-end latency of the $\MV$ ranges between $0.1895 ms$ and $0.6004 ms$. On the other hand, sweeping the 34 beams with exhaustive search and 802.11ad standard takes $1.27 ms$~\cite{Steinmetzer_2017}. Thus, we observe $52.72-85.07\%$ improvement in beam selection time compared to the 802.11ad standard.

\noindent
$\bullet$ {\bf Comparing with State-of-the-art Methods: }
In Tab.~\ref{tab:TableComp}, we compare the performance of the $\MV$ with the state-of-the-art mmWave beam selection methods by Klautau {\em et al.}~\cite{8642397}, Dias {\em et al.}~\cite{8815569}, and Xu {\em et al.}~\cite{9129762}. We report the average top-1 accuracy for LOS and NLOS scenarios. For the $\MV$, we use the accuracies from \third~twin and consider the 2dB threshold~($Acc_{(1,2)}$ in Eq.~\ref{eq:top-K-threshold}). We observe that the $\MV$ outperforms the competing methods by 6-46\% in accuracy. Unlike our work, other methods are only validated on synthetic data. The first two works use LiDAR sensors on the synthetically-generated Raymobtime dataset~\cite{8503086}, and Xu {\em et al.}~\cite{9129762} generate images using Blender software~\cite{Blender_website} and use them to construct a LiDAR-like point cloud.

\begin{table}[t!]
\centering
\resizebox{0.95\linewidth}{!}{
\begin{tabular}{||c|c|c|c|c||}
\hline
Papers & Top-$1$  &  Modalities & Method  & Validation \\ 
 & Acc.~(\%)   &  &  & data \\ \hline \hline
Klautau {\em et al.}~\cite{8642397}  & 30 & LiDAR & DL-based & Synthetic \\ \hline
Dias {\em et al.}~\cite{8815569} & 20 & LiDAR & DL-based  & Synthetic \\ \hline
Xu {\em et al.}~\cite{9129762} & 60 & Camera & DL-based & Synthetic  \\ \hline
{\begin{tabular}{@{}c@{}}
$\MV$ at the Edge\\ (this paper) 
    \end{tabular}}  & \textbf{66} &  {\begin{tabular}{@{}c@{}}
Camera, \\LiDAR, \\GPS
    \end{tabular}} & {\begin{tabular}{@{}c@{}}
DL-based\\ Digital twin
    \end{tabular}} & Real\\ \hline
\end{tabular}}
    \caption {Comparison of the proposed $\MV$ with the relevant state-of-the-art mmWave beam selection methods.}
   \label{tab:TableComp}
\end{table}

\section{Future Directions and Challenges}
\label{sec:future}

In this paper, we present a pioneering work on the digital twin field that offers different virtual world choices based on fidelity needs for mmWave beamforming. In the following, we provide future challenges and directions we plan on pursuing:

\noindent
$\bullet$ {\bf Computation Resources:} With ordinary computers today, ray tracing takes a long time to keep up with fast-pacing vehicular traffic. Thus, in today's standards expensive computing sources are needed. However, as the computation power for devices and the availability of such devices are in steady rise~\cite{10012285}, more powerful computations will be ubiquitous in the future, making ray tracing possible in the order $ms$.

\noindent
$\bullet$ {\bf Re-ray Tracing:} As indicated previously, ray tracing is the bottleneck of the digital twin in terms of computation and timing, because each time a new environment is entered, the ray tracing needs to be run again. In this paper, we propose addressing this by creating beam dictionaries by constantly running the twin so that when an appropriate beam is requested, the twin can provide an optimal answer. However, in case of an environment that the digital twin has not seen or generated before, ray tracing needs to be run from scratch. In order to cut the ray tracing time, we will develop more efficient computation strategies so that the ray tracing from scratch is not needed.

\noindent
$\bullet$ {\bf Reconfigurable Networks:} Currently, possible Tx beam options depend on preset beam patterns and directions, which brings suboptimal beam selections, necessitating alternative ways to deliver signals to the Rx. Given that real-time fast beam steering are at a development stage~\cite{FastBeamStrBas}, we will explore the computation feasibilities of beam generation that eliminates the dictionary dependency at Tx using digital twins. Additionally, reconfigurable intelligent surfaces~(RIS), which aims to (i) redirect and (ii) change the phase of incoming rays so that they constructively aggregate at a target Rx, gains attention for digital twins in recent years~\cite{hoydis2023sionna}, enabling reconfigurable networks. Accordingly, by constantly running the real world replica and updating the deployed ML models, the digital twins calculates optimal wireless parameters, such as standards, available bandwidth, beam features, beam directions, and RIS parameters. In future work, we will explore feasibility of reconfigurable networks using digital twins.

\noindent
$\bullet$ {\bf Computation Load and Data Granularity Trade-off:} As we discussed in Sec.\ref{sec:ValMultiverse} and \ref{sec:compareTwins}, accuracy and computation load in twins are inversely proportional. Additionally, the level of detail in multimodal dataset transmitted to digital twins affects the beam selection performance by altering the virtual world precision, hence ray tracing outcome. In future work, we will explore the impact of data granularity on the beam selection accuracy and computation load, for which we will investigate alternative methods to mitigate, such as transfer learning.


    

\section{Conclusions}
\label{sec:conclusion}
In this paper, we propose to expand the digital twin concept towards a $\MV$ of twins and demonstrate its application for mmWave beam selection in V2X scenarios. The $\MV$ play a vital role when local DL methods are insufficient due to environment changes, and the exhaustive beam search is difficult due to mobility. Each twin in the $\MV$, created in Wireless InSite simulator with different settings, captures the real world with different cost/fidelity trade-off. Accordingly, each twin prepares a beam selection dictionary for a quick reference when the environment change is detected. We validate the beam selection decisions by the $\MV$ through the experimental dataset, FLASH. Our evaluations show that this \emph{$\MV$ at the Edge} correctly predicts the top-$K$ beams with upto $85.22\%$ accuracy. Our $\MV$-based method yields upto $85.07\%$ improvement in beam selection time compared to 802.11ad standard.

\bibliographystyle{IEEEtran}
\bibliography{References}
\newpage
\newpage
\vfill

\end{document}